\documentclass[journal,10pt,twoside,twocolumn]{IEEEtran}

\usepackage{cite}
\usepackage[T1]{fontenc}
\usepackage{graphicx}
\usepackage{amssymb}
\usepackage{amsmath}
\usepackage{paralist}
\usepackage{subfigure}
\usepackage{setspace}
\usepackage{microtype}
\usepackage{hyperref}
\usepackage{url}
\usepackage{balance}
\usepackage{amsthm}
\usepackage{mathtools}
\usepackage{xcolor}

\usepackage{booktabs} 
\usepackage{algorithm}
\usepackage{algorithmic}
\usepackage{amsthm}
\usepackage{multirow}
\DeclareMathOperator\dif{d\!}
\usepackage{tablefootnote}

\usepackage{amssymb}
\usepackage{amsfonts}
\usepackage{mathrsfs}
\usepackage{xspace}
\usepackage{bm}
\usepackage{upgreek}

\newcommand{\safemath}[2]{\newcommand{#1}{\ensuremath{#2}\xspace}}

\safemath{\bma}{\mathbf{a}}
\safemath{\bmb}{\mathbf{b}}
\safemath{\bmc}{\mathbf{c}}
\safemath{\bmd}{\mathbf{d}}
\safemath{\bme}{\mathbf{e}}
\safemath{\bmf}{\mathbf{f}}
\safemath{\bmg}{\mathbf{g}}
\safemath{\bmh}{\mathbf{h}}
\safemath{\bmi}{\mathbf{i}}
\safemath{\bmj}{\mathbf{j}}
\safemath{\bmk}{\mathbf{k}}
\safemath{\bml}{\mathbf{l}}
\safemath{\bmm}{\mathbf{m}}
\safemath{\bmn}{\mathbf{n}}
\safemath{\bmo}{\mathbf{o}}
\safemath{\bmp}{\mathbf{p}}
\safemath{\bmq}{\mathbf{q}}
\safemath{\bmr}{\mathbf{r}}
\safemath{\bms}{\mathbf{s}}
\safemath{\bmt}{\mathbf{t}}
\safemath{\bmu}{\mathbf{u}}
\safemath{\bmv}{\mathbf{v}}
\safemath{\bmw}{\mathbf{w}}
\safemath{\bmx}{\mathbf{x}}
\safemath{\bmy}{\mathbf{y}}
\safemath{\bmz}{\mathbf{z}}
\safemath{\bmzero}{\mathbf{0}}
\safemath{\bmone}{\mathbf{1}}

\bmdefine{\biad}{a}
\bmdefine{\bibd}{b}
\bmdefine{\bicd}{c}
\bmdefine{\bidd}{d}
\bmdefine{\bied}{e}
\bmdefine{\bifd}{f}
\bmdefine{\bigd}{g}
\bmdefine{\bihd}{h}
\bmdefine{\biid}{i}
\bmdefine{\bijd}{j}
\bmdefine{\bikd}{k}
\bmdefine{\bild}{l}
\bmdefine{\bimd}{m}
\bmdefine{\bind}{n}
\bmdefine{\biod}{o}
\bmdefine{\bipd}{p}
\bmdefine{\biqd}{q}
\bmdefine{\bird}{r}
\bmdefine{\bisd}{s}
\bmdefine{\bitd}{t}
\bmdefine{\biud}{u}
\bmdefine{\bivd}{v}
\bmdefine{\biwd}{w}
\bmdefine{\bixd}{x}
\bmdefine{\biyd}{y}
\bmdefine{\bizd}{z}

\bmdefine{\bixid}{\xi}
\bmdefine{\bilambdad}{\lambda}
\bmdefine{\bimud}{\mu}
\bmdefine{\bithetad}{\theta}
\bmdefine{\biphid}{\phi}
\bmdefine{\bideltad}{\delta}

\safemath{\bmia}{\biad}
\safemath{\bmib}{\bibd}
\safemath{\bmic}{\bicd}
\safemath{\bmid}{\bidd}
\safemath{\bmie}{\bied}
\safemath{\bmif}{\bifd}
\safemath{\bmig}{\bigd}
\safemath{\bmih}{\bihd}
\safemath{\bmii}{\biid}
\safemath{\bmij}{\bijd}
\safemath{\bmik}{\bikd}
\safemath{\bmil}{\bild}
\safemath{\bmim}{\bimd}
\safemath{\bmin}{\bind}
\safemath{\bmio}{\biod}
\safemath{\bmip}{\bipd}
\safemath{\bmiq}{\biqd}
\safemath{\bmir}{\bird}
\safemath{\bmis}{\bisd}
\safemath{\bmit}{\bitd}
\safemath{\bmiu}{\biud}
\safemath{\bmiv}{\bivd}
\safemath{\bmiw}{\biwd}
\safemath{\bmix}{\bixd}
\safemath{\bmiy}{\biyd}
\safemath{\bmiz}{\bizd}

\safemath{\bmxi}{\bixid}
\safemath{\bmlambda}{\bilambdad}
\safemath{\bmmu}{\bimud}
\safemath{\bmtheta}{\bithetad}
\safemath{\bmphi}{\biphid}
\safemath{\bmdelta}{\bideltad}

\safemath{\bA}{\mathbf{A}}
\safemath{\bB}{\mathbf{B}}
\safemath{\bC}{\mathbf{C}}
\safemath{\bD}{\mathbf{D}}
\safemath{\bE}{\mathbf{E}}
\safemath{\bF}{\mathbf{F}}
\safemath{\bG}{\mathbf{G}}
\safemath{\bH}{\mathbf{H}}
\safemath{\bI}{\mathbf{I}}
\safemath{\bJ}{\mathbf{J}}
\safemath{\bK}{\mathbf{K}}
\safemath{\bL}{\mathbf{L}}
\safemath{\bM}{\mathbf{M}}
\safemath{\bN}{\mathbf{N}}
\safemath{\bO}{\mathbf{O}}
\safemath{\bP}{\mathbf{P}}
\safemath{\bQ}{\mathbf{Q}}
\safemath{\bR}{\mathbf{R}}
\safemath{\bS}{\mathbf{S}}
\safemath{\bT}{\mathbf{T}}
\safemath{\bU}{\mathbf{U}}
\safemath{\bV}{\mathbf{V}}
\safemath{\bW}{\mathbf{W}}
\safemath{\bX}{\mathbf{X}}
\safemath{\bY}{\mathbf{Y}}
\safemath{\bZ}{\mathbf{Z}}

\safemath{\bZero}{\mathbf{0}}
\safemath{\bOne}{\mathbf{1}}
\safemath{\bDelta}{\mathbf{\Delta}}
\safemath{\bLambda}{\mathbf{\UpLambda}}
\safemath{\bPhi}{\mathbf{\Upphi}}
\safemath{\bSigma}{\mathbf{\Upsigma}}
\safemath{\bOmega}{\mathbf{\Upomega}}
\safemath{\bTheta}{\mathbf{\Uptheta}}

\bmdefine{\biAd}{A}
\bmdefine{\biBd}{B}
\bmdefine{\biCd}{C}
\bmdefine{\biDd}{D}
\bmdefine{\biEd}{E}
\bmdefine{\biFd}{F}
\bmdefine{\biGd}{G}
\bmdefine{\biHd}{H}
\bmdefine{\biId}{I}
\bmdefine{\biJd}{J}
\bmdefine{\biKd}{K}
\bmdefine{\biLd}{L}
\bmdefine{\biMd}{M}
\bmdefine{\biOd}{N}
\bmdefine{\biPd}{O}
\bmdefine{\biQd}{P}
\bmdefine{\biRd}{R}
\bmdefine{\biSd}{S}
\bmdefine{\biTd}{T}
\bmdefine{\biUd}{U}
\bmdefine{\biVd}{V}
\bmdefine{\biWd}{W}
\bmdefine{\biXd}{X}
\bmdefine{\biYd}{Y}
\bmdefine{\biZd}{Z}

\bmdefine{\biDelta}{\Delta}
\bmdefine{\biLambda}{\Lambda}
\bmdefine{\biPhi}{\Phi}
\bmdefine{\biSigma}{\Sigma}
\bmdefine{\biOmega}{\Omega}
\bmdefine{\biTheta}{\Theta}

\safemath{\bimA}{\biAd}
\safemath{\bimB}{\biBd}
\safemath{\bimC}{\biCd}
\safemath{\bimD}{\biDd}
\safemath{\bimE}{\biEd}
\safemath{\bimF}{\biFd}
\safemath{\bimG}{\biGd}
\safemath{\bimH}{\biHd}
\safemath{\bimI}{\biId}
\safemath{\bimJ}{\biJd}
\safemath{\bimK}{\biKd}
\safemath{\bimL}{\biLd}
\safemath{\bimM}{\biMd}
\safemath{\bimN}{\biNd}
\safemath{\bimO}{\biOd}
\safemath{\bimP}{\biPd}
\safemath{\bimQ}{\biQd}
\safemath{\bimR}{\biRd}
\safemath{\bimS}{\biSd}
\safemath{\bimT}{\biTd}
\safemath{\bimU}{\biUd}
\safemath{\bimV}{\biVd}
\safemath{\bimW}{\biWd}
\safemath{\bimX}{\biXd}
\safemath{\bimY}{\biYd}
\safemath{\bimZ}{\biZd}

\safemath{\bimDelta}{\biDelta}
\safemath{\bimLambda}{\biLambda}
\safemath{\bimPhi}{\biPhi}
\safemath{\bimSigma}{\biSigma}
\safemath{\bimOmega}{\biOmega}
\safemath{\bimTheta}{\biTheta}

\safemath{\setA}{\mathcal{A}}
\safemath{\setB}{\mathcal{B}}
\safemath{\setC}{\mathcal{C}}
\safemath{\setD}{\mathcal{D}}
\safemath{\setE}{\mathcal{E}}
\safemath{\setF}{\mathcal{F}}
\safemath{\setG}{\mathcal{G}}
\safemath{\setH}{\mathcal{H}}
\safemath{\setI}{\mathcal{I}}
\safemath{\setJ}{\mathcal{J}}
\safemath{\setK}{\mathcal{K}}
\safemath{\setL}{\mathcal{L}}
\safemath{\setM}{\mathcal{M}}
\safemath{\setN}{\mathcal{N}}
\safemath{\setO}{\mathcal{O}}
\safemath{\setP}{\mathcal{P}}
\safemath{\setQ}{\mathcal{Q}}
\safemath{\setR}{\mathcal{R}}
\safemath{\setS}{\mathcal{S}}
\safemath{\setT}{\mathcal{T}}
\safemath{\setU}{\mathcal{U}}
\safemath{\setV}{\mathcal{V}}
\safemath{\setW}{\mathcal{W}}
\safemath{\setX}{\mathcal{X}}
\safemath{\setY}{\mathcal{Y}}
\safemath{\setZ}{\mathcal{Z}}
\safemath{\emptySet}{\varnothing}

\safemath{\colA}{\mathscr{A}}
\safemath{\colB}{\mathscr{B}}
\safemath{\colC}{\mathscr{C}}
\safemath{\colD}{\mathscr{D}}
\safemath{\colE}{\mathscr{E}}
\safemath{\colF}{\mathscr{F}}
\safemath{\colG}{\mathscr{G}}
\safemath{\colH}{\mathscr{H}}
\safemath{\colI}{\mathscr{I}}
\safemath{\colJ}{\mathscr{J}}
\safemath{\colK}{\mathscr{K}}
\safemath{\colL}{\mathscr{L}}
\safemath{\colM}{\mathscr{M}}
\safemath{\colN}{\mathscr{N}}
\safemath{\colO}{\mathscr{O}}
\safemath{\colP}{\mathscr{P}}
\safemath{\colQ}{\mathscr{Q}}
\safemath{\colR}{\mathscr{R}}
\safemath{\colS}{\mathscr{S}}
\safemath{\colT}{\mathscr{T}}
\safemath{\colU}{\mathscr{U}}
\safemath{\colV}{\mathscr{V}}
\safemath{\colW}{\mathscr{W}}
\safemath{\colX}{\mathscr{X}}
\safemath{\colY}{\mathscr{Y}}
\safemath{\colZ}{\mathscr{Z}}

\safemath{\opA}{\mathbb{A}}
\safemath{\opB}{\mathbb{B}}
\safemath{\opC}{\mathbb{C}}
\safemath{\opD}{\mathbb{D}}
\safemath{\opE}{\mathbb{E}}
\safemath{\opF}{\mathbb{F}}
\safemath{\opG}{\mathbb{G}}
\safemath{\opH}{\mathbb{H}}
\safemath{\opI}{\mathbb{I}}
\safemath{\opJ}{\mathbb{J}}
\safemath{\opK}{\mathbb{K}}
\safemath{\opL}{\mathbb{L}}
\safemath{\opM}{\mathbb{M}}
\safemath{\opN}{\mathbb{N}}
\safemath{\opO}{\mathbb{O}}
\safemath{\opP}{\mathbb{P}}
\safemath{\opQ}{\mathbb{Q}}
\safemath{\opR}{\mathbb{R}}
\safemath{\opS}{\mathbb{S}}
\safemath{\opT}{\mathbb{T}}
\safemath{\opU}{\mathbb{U}}
\safemath{\opV}{\mathbb{V}}
\safemath{\opW}{\mathbb{W}}
\safemath{\opX}{\mathbb{X}}
\safemath{\opY}{\mathbb{Y}}
\safemath{\opZ}{\mathbb{Z}}
\safemath{\opZero}{\mathbb{O}}
\safemath{\identityop}{\opI}

\safemath{\veca}{\bma}
\safemath{\vecb}{\bmb}
\safemath{\vecc}{\bmc}
\safemath{\vecd}{\bmd}
\safemath{\vece}{\bme}
\safemath{\vecf}{\bmf}
\safemath{\vecg}{\bmg}
\safemath{\vech}{\bmh}
\safemath{\veci}{\bmi}
\safemath{\vecj}{\bmj}
\safemath{\veck}{\bmk}
\safemath{\vecl}{\bml}
\safemath{\vecm}{\bmm}
\safemath{\vecn}{\bmn}
\safemath{\veco}{\bmo}
\safemath{\vecp}{\bmp}
\safemath{\vecq}{\bmq}
\safemath{\vecr}{\bmr}
\safemath{\vecs}{\bms}
\safemath{\vect}{\bmt}
\safemath{\vecu}{\bmu}
\safemath{\vecv}{\bmv}
\safemath{\vecw}{\bmw}
\safemath{\vecx}{\bmx}
\safemath{\vecy}{\bmy}
\safemath{\vecz}{\bmz}

\safemath{\veczero}{\bmzero}
\safemath{\vecone}{\bmone}
\safemath{\vecxi}{\bmxi}
\safemath{\veclambda}{\bmlambda}
\safemath{\vecmu}{\bmmu}
\safemath{\vectheta}{\bmtheta}
\safemath{\vecphi}{\bmphi}
\safemath{\vecdelta}{\bmdelta}

\safemath{\matA}{\bA}
\safemath{\matB}{\bB}
\safemath{\matC}{\bC}
\safemath{\matD}{\bD}
\safemath{\matE}{\bE}
\safemath{\matF}{\bF}
\safemath{\matG}{\bG}
\safemath{\matH}{\bH}
\safemath{\matI}{\bI}
\safemath{\matJ}{\bJ}
\safemath{\matK}{\bK}
\safemath{\matL}{\bL}
\safemath{\matM}{\bM}
\safemath{\matN}{\bN}
\safemath{\matO}{\bO}
\safemath{\matP}{\bP}
\safemath{\matQ}{\bQ}
\safemath{\matR}{\bR}
\safemath{\matS}{\bS}
\safemath{\matT}{\bT}
\safemath{\matU}{\bU}
\safemath{\matV}{\bV}
\safemath{\matW}{\bW}
\safemath{\matX}{\bX}
\safemath{\matY}{\bY}
\safemath{\matZ}{\bZ}
\safemath{\matzero}{\bmzero}

\safemath{\matDelta}{\bDelta}
\safemath{\matLambda}{\bLambda}
\safemath{\matPhi}{\bPhi}
\safemath{\matSigma}{\bSigma}
\safemath{\matOmega}{\bOmega}
\safemath{\matTheta}{\bTheta}

\safemath{\matidentity}{\matI}
\safemath{\matone}{\matO}

\safemath{\rnda}{A}
\safemath{\rndb}{B}
\safemath{\rndc}{C}
\safemath{\rndd}{D}
\safemath{\rnde}{E}
\safemath{\rndf}{F}
\safemath{\rndg}{G}
\safemath{\rndh}{H}
\safemath{\rndi}{I}
\safemath{\rndj}{J}
\safemath{\rndk}{K}
\safemath{\rndl}{L}
\safemath{\rndm}{M}
\safemath{\rndn}{N}
\safemath{\rndo}{O}
\safemath{\rndp}{P}
\safemath{\rndq}{Q}
\safemath{\rndr}{R}
\safemath{\rnds}{S}
\safemath{\rndt}{T}
\safemath{\rndu}{U}
\safemath{\rndv}{V}
\safemath{\rndw}{W}
\safemath{\rndx}{X}
\safemath{\rndy}{Y}
\safemath{\rndz}{Z}

\safemath{\rveca}{\bimA}
\safemath{\rvecb}{\bimB}
\safemath{\rvecc}{\bimC}
\safemath{\rvecd}{\bimD}
\safemath{\rvece}{\bimE}
\safemath{\rvecf}{\bimF}
\safemath{\rvecg}{\bimG}
\safemath{\rvech}{\bimH}
\safemath{\rveci}{\bimI}
\safemath{\rvecj}{\bimJ}
\safemath{\rveck}{\bimK}
\safemath{\rvecl}{\bimL}
\safemath{\rvecm}{\bimM}
\safemath{\rvecn}{\bimN}
\safemath{\rveco}{\bomO}
\safemath{\rvecp}{\bimP}
\safemath{\rvecq}{\bimQ}
\safemath{\rvecr}{\bimR}
\safemath{\rvecs}{\bimS}
\safemath{\rvect}{\bimT}
\safemath{\rvecu}{\bimU}
\safemath{\rvecv}{\bimV}
\safemath{\rvecw}{\bimW}
\safemath{\rvecx}{\bimX}
\safemath{\rvecy}{\bimY}
\safemath{\rvecz}{\bimZ}

\safemath{\rvecxi}{\bmxi}
\safemath{\rveclambda}{\bmlambda}
\safemath{\rvecmu}{\bmmu}
\safemath{\rvectheta}{\bmtheta}
\safemath{\rvecphi}{\bmphi}

\safemath{\rmatA}{\bimA}
\safemath{\rmatB}{\bimB}
\safemath{\rmatC}{\bimC}
\safemath{\rmatD}{\bimD}
\safemath{\rmatE}{\bimE}
\safemath{\rmatF}{\bimF}
\safemath{\rmatG}{\bimG}
\safemath{\rmatH}{\bimH}
\safemath{\rmatI}{\bimI}
\safemath{\rmatJ}{\bimJ}
\safemath{\rmatK}{\bimK}
\safemath{\rmatL}{\bimL}
\safemath{\rmatM}{\bimM}
\safemath{\rmatN}{\bimN}
\safemath{\rmatO}{\bimO}
\safemath{\rmatP}{\bimP}
\safemath{\rmatQ}{\bimQ}
\safemath{\rmatR}{\bimR}
\safemath{\rmatS}{\bimS}
\safemath{\rmatT}{\bimT}
\safemath{\rmatU}{\bimU}
\safemath{\rmatV}{\bimV}
\safemath{\rmatW}{\bimW}
\safemath{\rmatX}{\bimX}
\safemath{\rmatY}{\bimY}
\safemath{\rmatZ}{\bimZ}

\safemath{\rmatDelta}{\bimDelta}
\safemath{\rmatLambda}{\bimLambda}
\safemath{\rmatPhi}{\bimPhi}
\safemath{\rmatSigma}{\bimSigma}
\safemath{\rmatOmega}{\bimOmega}
\safemath{\rmatTheta}{\bimTheta}

\usepackage{amssymb}
\usepackage{amsfonts}
\usepackage{mathrsfs}
\usepackage{xspace}
\usepackage{bm}
\usepackage{fancyref}
\usepackage{textcomp}

\usepackage{multirow}
\usepackage{stmaryrd}

\newenvironment{textbmatrix}{	\setlength{\arraycolsep}{2.5pt}%
								\big[\begin{matrix}}{\end{matrix}\big]%
								\raisebox{0.08ex}{\vphantom{M}}}

\def\be{\begin{equation}}
\def\ee{\end{equation}}
\def\een{\nonumber \end{equation}}
\def\mat{\begin{bmatrix}}
\def\emat{\end{bmatrix}}
\def\btm{\begin{textbmatrix}}
\def\etm{\end{textbmatrix}}

\def\ba#1\ea{\begin{align}#1\end{align}}
\def\bas#1\eas{\begin{align*}#1\end{align*}}
\def\bs#1\es{\begin{split}#1\end{split}} 
\def\bg#1\eg{\begin{gather}#1\end{gather}}
\def\bml#1\eml{\begin{multline}#1\end{multline}}
\def\bi#1\ei{\begin{itemize}#1\end{itemize}}

\newcommand{\lefto}{\mathopen{}\left}

\DeclareMathOperator*{\argmin}{arg\;min}		

\newcommand{\abs}[1]{\lefto\lvert#1\right\rvert}		

\safemath{\dirac}{\delta}					
\safemath{\krond}{\dirac}					

\safemath{\upto}{\uparrow}
\safemath{\downto}{\downarrow}
\safemath{\iu}{j}							
\safemath{\ev}{\lambda}						
\safemath{\hilseqspace}{l^{2}}				
\newcommand{\banachfunspace}[1]{\setL^{#1}}	
\safemath{\hilfunspace}{\banachfunspace{2}}	

\safemath{\SNR}{\textsf{SNR}} 				
\safemath{\PAR}{\textsf{PAR}} 				
\safemath{\No}{N_0}							
\safemath{\Es}{E_s}							
\safemath{\Eb}{E_b}							
\safemath{\EbNo}{\frac{\Eb}{\No}}
\safemath{\EsNo}{\frac{\Es}{\No}}

\DeclareMathOperator{\CHop}{\ensuremath{\opH}} 
\safemath{\tvir}{\rndh_{\CHop}}				
\safemath{\tvtf}{\rndl_{\CHop}}				
\safemath{\spf}{\rnds_{\CHop}}				
\safemath{\bff}{H_{\CHop}}					

\safemath{\ircf}{r_{h}}						
\safemath{\tftvcf}{r_{s}}					
\safemath{\tfcf}{r_{l}}						
\safemath{\bfcf}{r_{H}}						

\safemath{\tcorr}{c_h}						
\safemath{\scf}{c_{s}}						
\safemath{\tfcorr}{c_{l}}					
\safemath{\fcorr}{c_{H}}						

\safemath{\mi}{I}							
\safemath{\capacity}{C}						

\safemath{\normal}{\mathcal{N}}			
\safemath{\jpg}{\mathcal{CN}}			
\safemath{\mchain}{\leftrightarrow}		

\safemath{\dB}{\,\mathrm{dB}}
\safemath{\dBm}{\,\mathrm{dBm}}
\safemath{\Hz}{\,\mathrm{Hz}}
\safemath{\kHz}{\,\mathrm{kHz}}
\safemath{\MHz}{\,\mathrm{MHz}}
\safemath{\GHz}{\,\mathrm{GHz}}
\safemath{\s}{\,\mathrm{s}}
\safemath{\ms}{\,\mathrm{ms}}
\safemath{\mus}{\,\mathrm{\text{\textmu}s}}
\safemath{\ns}{\,\mathrm{ns}}
\safemath{\ps}{\,\mathrm{ps}}
\safemath{\meter}{\,\mathrm{m}}
\safemath{\mm}{\,\mathrm{mm}}
\safemath{\cm}{\,\mathrm{cm}}
\safemath{\m}{\,\mathrm{m}}
\safemath{\W}{\,\mathrm{W}}
\safemath{\mW}{\, \mathrm{mW}}
\safemath{\J}{\,\mathrm{J}}
\safemath{\K}{\,\mathrm{K}}
\safemath{\bit}{\,\mathrm{bit}}
\safemath{\nat}{\,\mathrm{nat}}

\safemath{\define}{\triangleq}			

\safemath{\equivalent}{\sim}
\safemath{\distas}{\sim}					
\safemath{\sdiff}{\Delta}				

\safemath{\reals}{\mathbb{R}}
\safemath{\positivereals}{\reals_{+}}
\safemath{\integers}{\mathbb{Z}}
\safemath{\posint}{\integers_{+}}
\safemath{\naturals}{\mathbb{N}}
\safemath{\posnaturals}{\naturals_{+}}
\safemath{\complexset}{\mathbb{C}}
\safemath{\rationals}{\mathbb{Q}}

\newcommand*{\fancyrefapplabelprefix}{app}		
\newcommand*{\fancyrefthmlabelprefix}{thm}		
\newcommand*{\fancyreflemlabelprefix}{lem}		
\newcommand*{\fancyrefcorlabelprefix}{cor}		
\newcommand*{\fancyrefdeflabelprefix}{def}		
\newcommand*{\fancyrefproplabelprefix}{prop}		
\newcommand*{\fancyrefobslabelprefix}{obs}		
\newcommand*{\fancyrefexmpllabelprefix}{exmpl}
\newcommand*{\fancyrefalglabelprefix}{alg}		

\frefformat{vario}{\fancyrefseclabelprefix}{Section~#1}
\frefformat{vario}{\fancyrefthmlabelprefix}{Theorem~#1}
\frefformat{vario}{\fancyreflemlabelprefix}{Lemma~#1}
\frefformat{vario}{\fancyrefcorlabelprefix}{Corollary~#1}
\frefformat{vario}{\fancyrefdeflabelprefix}{Definition~#1}
\frefformat{vario}{\fancyrefobslabelprefix}{Observation~#1}
\frefformat{vario}{\fancyreffiglabelprefix}{Fig.~#1}
\frefformat{vario}{\fancyrefapplabelprefix}{Appendix~#1} 
\frefformat{vario}{\fancyrefeqlabelprefix}{(#1)}
\frefformat{vario}{\fancyrefproplabelprefix}{Proposition~#1}
\frefformat{vario}{\fancyrefexmpllabelprefix}{Example~#1}
\frefformat{vario}{\fancyrefalglabelprefix}{Algorithm~#1}

\newtheorem{thm}{Theorem}

\newtheorem{defi}{Definition}
\newtheorem{lem}{Lemma} 

\safemath{\dictab}{[\,\dicta\,\,\dictb\,]}

\safemath{\ysig}{\bmy}
\safemath{\ysighat}{\hat{\ysig}}
\safemath{\ysigdim}{M}
\safemath{\xsig}{\bmx}
\safemath{\xsigdim}{N}
\safemath{\nx}{n_x}
\safemath{\zsig}{\bmz}
\safemath{\zsigdim}{\ysigdim}
\safemath{\rsig}{\bmr}
\safemath{\Adict}{\bA}
\safemath{\Adicttilde}{\widetilde{\Adict}}
\safemath{\Adictdim}{\outputdim\times\xsigdim}
\safemath{\avec}{\bma}
\safemath{\avectilde}{\tilde{\avec}}
\safemath{\Bdict}{\bB}
\safemath{\Bdicttilde}{\widetilde{\Bdict}}
\safemath{\Cdict}{\bC}
\safemath{\cvec}{\bmc}
\safemath{\Ddict}{\bD}
\safemath{\Ddictdim}{\ysigdim\times\xsigdim}
\safemath{\dvec}{\bmd}
\safemath{\Ddicttilde}{\widetilde{\bD}}
\safemath{\Bonb}{\bB}
\safemath{\bvec}{\bmb}
\safemath{\Bonbdim}{\ysigdim\times\ysigdim}
\safemath{\noise}{\bmn}
\safemath{\noisedim}{\ysigim}
\safemath{\err}{\bme}
\safemath{\errdim}{\ysigdim}
\safemath{\errset}{\setE}
\safemath{\nerr}{n_e}
\safemath{\delop}{\bP_\errset}
\safemath{\delopc}{\bP_{{\errset}^c}}

\safemath{\cplxi}{\imath}
\safemath{\cplxj}{\jmath}

\safemath{\dict}{\matD}
\safemath{\inputdim}{N}		
\safemath{\outputdim}{M}		
\safemath{\sparsity}{S}	
\safemath{\inputdimA}{{N_a}}	
\safemath{\inputdimB}{{N_b}}	
\safemath{\elemA}{{n_a}}	
\safemath{\elemB}{{n_b}}	
\safemath{\resA}{\matR_a}	
\safemath{\resB}{\matR_b}	
\safemath{\subD}{\matS} 
\safemath{\subA}{\matS_a} 
\safemath{\subB}{\matS_b} 
\safemath{\dicta}{\matA} 	
\safemath{\dictb}{\matB} 	
\safemath{\hollowS}{H}
\safemath{\hollowA}{H_a}
\safemath{\hollowB}{H_b}
\safemath{\cross}{Z}
\safemath{\coh}{\mu_d}			
\safemath{\coha}{\mu_a}			
\safemath{\cohb}{\mu_b}			
\safemath{\mubs}{\nu}	
\safemath{\cohm}{\mu_m} 
\safemath{\dictset}{\setD}	
\safemath{\dictsetp}{\dictset(\coh,\coha,\cohb)}	
\safemath{\dictsetgen}{\dictset_\text{gen}}
\safemath{\dictsetgenp}{\dictsetgen(\coh)}
\safemath{\dictsetonb}{\dictset_\text{onb}}
\safemath{\dictsetonbp}{\dictsetonb(\coh)}

\safemath{\leftside}{U}
\safemath{\rightsideA}{R_a}
\safemath{\rightsideB}{R_b}

\safemath{\indexS}{\setI_S} 

\safemath{\na}{n_a}			
\safemath{\nb}{n_b}			
\safemath{\coeffa}{p_i}	
\safemath{\coeffb}{q_j}	
\safemath{\seta}{\setP}		
\safemath{\setb}{\setQ}     
\safemath{\setw}{\setW}	
\safemath{\setz}{\setZ}	
\safemath{\cola}{\veca}		
\safemath{\colb}{\vecb}		
\safemath{\cold}{\vecd}		
\safemath{\inputvec}{\vecx} 	
\safemath{\error}{\vece}	
\safemath{\noiseout}{\vecz} 	
\safemath{\inputvecel}{x}
\safemath{\inputveca}{\vecx_a}
\safemath{\inputvecb}{\vecx_b}
\safemath{\outputvec}{\vecy}	
\safemath{\lambdamin}{\lambda_{\mathrm{min}}}

\safemath{\elltwo}{\ell_2}
\safemath{\ellone}{\ell_1}
\safemath{\ellzero}{\ell_0}
\safemath{\ellinf}{\ell_\infty}
\safemath{\ellinftilde}{\ell_{\widetilde\infty}}
\safemath{\licard}{Z(\coh,\coha,\cohb)}
\safemath{\xsol}{\hat{x}}
\safemath{\xbord}{x_b}		
\safemath{\xstat}{x_s}		
\safemath{\xstatLone}{\tilde{x}_s}
\safemath{\order}{\mathcal{O}} 
\safemath{\scales}{\Theta} 
\safemath{\ones}{\mathbf{1}} 
\safemath{\zeroes}{\mathbf{0}} 
\safemath{\thlone}{\kappa(\coh,\cohb)} 
\safemath{\constoneA}{\delta} 
\safemath{\constoneB}{\epsilon} 
\safemath{\nlarge}{L}				   
\safemath{\sumlarge}{S_\nlarge}
\safemath{\maxlarger}{P_\nlarge}	   
\safemath{\Pzero}{\textrm{P0}}	
\safemath{\Pone}{\textrm{P1}}
\safemath{\vecfir}{\vecw}			 
\safemath{\vecsec}{\vecz}
\safemath{\elvecfir}{w}              
\safemath{\elvecsec}{z}				 
\safemath{\nlargefir}{n}
\safemath{\normout}{\gamma}
\safemath{\auxfun}{h}
\safemath{\supp}{\textrm{supp}}

\safemath{\indexa}{\ell}
\safemath{\indexb}{r}
\safemath{\indexc}{i}
\safemath{\indexd}{j}

\safemath{\project}{P}

\safemath{\MSEzero}{\mathsf{0\textnormal{-}MSE}}
\safemath{\MSEzeroIJT}{\mathsf{0\textnormal{-}MSE}^{(m,n)}_t}
\safemath{\MSEzeroIJomega}{\mathsf{0\textnormal{-}MSE}^{(m,n)}_\omega}
\safemath{\MSEone}{\mathsf{1\textnormal{-}MSE}}
\safemath{\MSEoneIJT}{\mathsf{1\textnormal{-}MSE}^{(m,n)}_t}

\safemath{\bHfreq}{{\bH}_\omega}
\safemath{\bHfreqPerfectCSI}{{\bH}_\omega^\textnormal{CSI}}
\safemath{\bHfreqIJ}{[\bHfreq]_{m,n}}

\safemath{\bHtimeuncor}{\widehat{\bH}_\ell^\textnormal{uncor}}
\safemath{\bHtime}{\widehat{\bH}_\ell}
\safemath{\bHtimePerfectCSI}{\bHtime^\textnormal{CSI}}
\safemath{\bHtimeIJPerfectCSI}{[\bHtimePerfectCSI]_{m,n}}
\safemath{\bHtimenoidx}{\widehat{\bH}}
\safemath{\bHtimeprime}{\widehat{\bH}_{\ell'}}
\safemath{\HtimeIJ}{[\bHtime]_{m,n}}

\safemath{\bGfreq}{{\bG}_\omega}
\safemath{\bGfreqinterp}{\widetilde{\bG}_\omega}

\safemath{\dmax}{d_{\textnormal{max}}}

\safemath{\bpset}{{\vert\setP\vert}}
\safemath{\activeset}{{\vert\Omega\vert}}

\newcommand{\first}{1st }
\newcommand{\zeroth}{0th }

\newcommand{\revision}[1]{\textcolor{black}{#1}}
\newcommand{\revisiontwo}[1]{\textcolor{black}{#1}}

\IEEEoverridecommandlockouts 
\allowdisplaybreaks 

\usepackage{color}

\newtheorem{corollary}{Corollary}

\begin{document}

\title{Approximate Gram-Matrix Interpolation \\ for Wideband Massive MU-MIMO Systems}

\author{ 
Charles Jeon, Zequn Li, and Christoph Studer
\thanks{C.~Jeon and C.~Studer are with the School of Electrical and Computer Engineering, Cornell University, Ithaca, NY, USA; e-mail: jeon@csl.cornell.edu, studer@cornell.edu; website: \url{vip.ece.cornell.edu}}
\thanks{Z.~Li was a visiting researcher at the School of Electrical and Computer Engineering, Cornell University, Ithaca, NY and is now with Department of Automation, Tsinghua University, Beijing, China; e-mail: li-zq13@mails.tsinghua.edu.cn}
\thanks{A simulation framework to reproduce the results of this paper will be made available on GitHub after (possible) acceptance.}
}

\maketitle

\begin{abstract}
\revision{Numerous linear and non-linear data-detection and precoding algorithms for wideband massive multi-user (MU) multiple-input multiple-output (MIMO) wireless systems that rely on orthogonal frequency-division multiplexing~(OFDM) or single-carrier frequency-division multiple access (SC-FDMA) require the computation of the Gram matrix for each active subcarrier. 
However, computing the Gram matrix for each active subcarrier results in excessively high computational complexity. 
In this paper, we propose novel, approximate algorithms that significantly reduce the complexity of Gram-matrix computation  by simultaneously exploiting correlation across subcarriers and channel hardening.
We  show analytically that a small fraction of Gram-matrix computations in combination with approximate interpolation schemes are sufficient to achieve near-optimal error-rate performance at low computational complexity in massive MU-MIMO systems.
We furthermore demonstrate  the improved robustness of our methods against channel-estimation errors compared to exact Gram-matrix interpolation algorithms that typically require high computational complexity.}
\end{abstract}






\section{Introduction}
\label{sec:introduction}
\IEEEPARstart{M}{ASSIVE} multi-user (MU) multiple-input multiple-output (MIMO) will be a key technology in fifth-generation (5G) wireless systems~\cite{LETM2014,Marzetta10}.
The idea of massive MU-MIMO is to equip the base-station (BS) with hundreds  of antenna elements while serving tens of user equipments (UEs) in the same time-frequency resource. Such large antenna arrays enable extremely fine-grained beamforming in the uplink (UEs transmit to the BS) and in the downlink (BS transmits to the UEs), which offers superior spectral efficiency compared to traditional, small-scale MIMO technology that use only a few antennas at the BS.

In the uplink, linear data-detection algorithms that rely on minimum-mean square error (MMSE) equalization or zero-forcing (ZF) equalization are known to achieve near-optimal error-rate performance in realistic massive MU-MIMO systems with a finite number of transmit antennas \cite{HBD11,paulraj03,LETM2014}.
Non-linear data-detection algorithms  \cite{indiachemp,JGMS2015conf,JMS2016} have recently been shown to outperform linear methods in systems where number of UEs is comparable to the number of BS antennas.
\revision{Most of these linear and non-linear data-detection algorithms entail high computational complexity, often dominated by the computation of the so-called \emph{Gram matrix} $\bG=\bH^H\bH$~\cite{WYWDCS2014,BeiYinThesis}. Here, $\bH\in\complexset^{B\times U}$ is the (uplink) channel matrix, $B$ is the number of BS antennas, and $U$ is the number of (single-antenna) users.}
The computational complexity is orders-of-magnitude higher in wideband systems that use orthogonal frequency-division multiplexing (OFDM) or single-carrier frequency-division multiple access (SC-FDMA), in which a Gram matrix must be computed for each active subcarrier (i.e., subcarriers used for pilots or data transmission)~\cite{WYWDCS2014}.
\revisiontwo{For example, Gram matrix computation requires more than $2\times$ higher complexity than data detection for a $128$ BS antenna $16$ UE antenna MU-MIMO system \cite[Table 4.2]{BeiYinThesis}, and is much higher for a system with more BS antennas}. 
\revision{In the massive MU-MIMO downlink, precoding is necessary to focus the transmit energy towards the UEs and to mitigate multi-user interference~\cite{LETM2014}. In wideband systems, the  complexity of linear precoding algorithms is---analogously to the uplink---dominated by Gram matrix computation  on all active subcarriers.}

While some data-detection and precoding algorithms have been proposed that avoid the computation of the Gram matrix altogether (see, e.g., \cite{yin2014conjugate,wu2016efficient,WDCS2016}), these methods do not allow the re-use of intermediate results in time-division duplexing (TDD) systems. Specifically, the Gram matrix and its inverse cannot be re-used in the uplink (for equalization) and downlink (for precoding), which would significantly lower the computational complexity. Hence, such algorithms inevitably perform redundant computations during data-detection and precoding, which leads to inefficient transceiver designs. 


%
%
%
%
%
%
\subsection{Interpolation-Based Matrix Computations}
In practical wideband communication systems, e.g., building upon IEEE 802.11n~\cite{IEEE11n} and 3GPP-LTE~\cite{3gpp36.211}, the channel's delay spread is often substantially smaller than the number of active subcarriers.
Hence, the channel coefficients are correlated across subcarriers. This property can be exploited to reduce the computational complexity of commonly-used matrix computations required in multi-antenna systems.
\revision{More specifically, the papers \cite{CBBHB2005,BB2004,CB2011,CB2010} avoid a brute-force approach in traditional, small-scale, and point-to-point MIMO-OFDM systems by using exact interpolation-based algorithms for matrix inversion and QR factorization.}
While a few hardware designs~\cite{CCH2011,CHCH2011} have demonstrated the efficacy of these exact interpolation methods in small-scale MIMO systems, their complexity does not scale well to wideband massive MU-MIMO systems with hundreds  of BS antennas, tens of users, and thousands of subcarriers;
\revisiontwo{
in fact, recent 3GPP specifications on New Radio (NR) access technology shows that the number of active subcarriers is 3300 or 6600 in Rel-15 \cite{3gpp38912,J2018}.
}
%
\revision{In addition, the impact of imperfect channel-state information (CSI) and antenna correlation on such exact, interpolation-based matrix computation algorithms is routinely ignored, but significantly affects their performance in practical scenarios (see \fref{sec:numericalresults} for a detailed discussion).}

\subsection{Contributions} 
\revision{Inspired by exact, interpolation-based matrix computation algorithms in \cite{CBBHB2005,BB2004,CB2011,CB2010} for  small-scale wideband MIMO systems, we propose novel algorithms for \emph{approximate} Gram matrix computation in massive MU-MIMO systems.}
We start by establishing the minimum number of Gram matrix base-points that are required for exact interpolation. 
We then show that channel-hardening in massive MU-MIMO enables approximate interpolation schemes that achieve near-exact \revision{error-rate performance}, even with strong undersampling in the frequency domain.
In particular, we provide analytical results that characterize the approximation errors of the proposed interpolation methods depending on the channel's delay spread and the antenna configuration.
\revision{
We furthermore derive exact mean-squared error (MSE) expressions of our approximate interpolation algorithms for imperfect CSI and BS-antenna correlation.
We characterize the trade-offs between computational complexity and error-rate performance in realistic massive MU-MIMO-OFDM systems, and we demonstrate the robustness of our approximate interpolation methods for realistic scenarios with imperfect CSI and BS-antenna correlation.}

\subsection{\revision{Relevant Prior Art}}

\revision{
Data detection and precoding for small-scale, single- and multi-carrier MIMO systems is a well studied topic; see e.g. \cite{SP02,SSBGS02,SSP01,paulraj03,studer2011asic} and the references therein. 
\revisiontwo{
Data detection and precoding algorithms for massive MIMO systems have been proposed in, e.g.,  \cite{CL2014,JGMS2015conf,CLSK2013,ML12a,WYWDCS2014,indiachemp}, which leverage the fact that the Gram matrix is diagonally dominant  \cite{RPLLMET13}.
}
However, all of these results 
(i) do not exploit specific properties of massive MU-MIMO systems and (ii) ignore the fact that time-division duplexing (TDD)-based systems must perform data-detection \emph{and} precoding, and hence, can re-use intermediate results (such as the Gram matrix) to reduce the computational complexity.
In contrast, our results exploit the specifics of massive MU-MIMO systems, namely channel hardening, and enable a re-use  of the computations carried out in the uplink for downlink precoding.}

\revision{
The recent report \cite{MAMMOET} proposed an approximate interpolation-based ZF-based equalizer for a wideband massive MU-MIMO testbed. In contrast to our work, the authors interpolate the \emph{inverse} of the Gram matrix. While simulation results in \cite{MAMMOET} show that the method works well in practice, no theoretical results have been provided. 
In contrast, we use approximate methods to interpolate the Gram matrix, and we provide exact analytical results that provide a solid foundation of approximate interpolation methods in massive MU-MIMO systems. 
}

\subsection{Notation} 
Lowercase and uppercase boldface letters stand column vectors and matrices, respectively. The transpose, Hermitian, and pseudo-inverse of the matrix $\bA$ are denoted by $\bA^T$, $\bA^H$, and $\bA^{\dag}$, respectively. 
We use $[\bA]_{m,n}$ to represent the  $m$th row and $n$th column entry of the matrix $\bA$. 
%
Sets are designated by uppercase calligraphic letters, and $\vert\setA\vert$ denotes the cardinality of the set $\setA$. The complex conjugate of $a\in\complexset$ is $a^*$.
\revision{
The indicator function is defined as $\mathbb{I}(a)$, where $\mathbb{I}(a)=1$ if $a$ is true and $0$ otherwise.  
$\setC\setN(\bm\mu,\bm\Sigma)$ designates the complex Gaussian distribution with mean vector $\bm\mu$ and covariance matrix $\bm\Sigma$.
}

\subsection{Paper Outline} 
The rest of the paper is organized as follows. \fref{sec:prereq} introduces the necessary prerequisites. \fref{sec:grammatrixcomputation} proposes our exact and approximate interpolation-based Gram matrix computation algorithms.  \fref{sec:errorbound} and \fref{sec:complexityanalysis} provide an analysis of the approximation error  and complexity for the proposed approximate interpolation methods, respectively. \fref{sec:numericalresults} shows numerical results. We conclude in \fref{sec:conclusions}.


\section{Prerequisites}\label{sec:prereq}

\revision{We start by summarizing the considered wideband massive MU-MIMO system and channel model. We then outline computationally-efficient ways for linear data detection and precoding that make use of the Gram matrix.}

\subsection{System Model}
\label{sec:systemmodel}

Without loss of generality, we focus on the uplink\footnote{By assuming channel reciprocity~\cite{LETM2014}, our results directly apply to the downlink, in which $\bH^T$ is the downlink channel matrix. Once all of the Gram matrices have been computed, they can be re-used in the downlink for linear (e.g., zero-forcing or Wiener filter) precoding. See \fref{sec:detandpre} for the details.} of a wideband massive MU-MIMO system with $B$ base-station antennas, $U$ single-antenna UEs (with $U\ll B$), and~$W$ subcarriers.
For each active subcarrier $\omega\in\Omega$ with $\Omega$ containing the indices of the active (data and pilot) subcarriers, we model the received frequency-domain (FD) signal as follows:
\begin{align} \label{eq:inputoutput}
\bmy_\omega = \bHfreq \bms_\omega+ \bmn_\omega.
\end{align} 
Here,~$\bHfreq\in\complexset^{B\times U}$ is the FD channel matrix, $\bms_\omega\in\complexset^U$ is the transmit vector, and $\bmn_\omega\in\complexset^B$ models additive noise. 
The FD input-output relation in \fref{eq:inputoutput} is able to model both OFDM and SC-FDMA systems. For OFDM systems, the entries of the transmit vector~$\bms_\omega$ are taken from a discrete constellation set $\setO$ (e.g., 16-QAM); in SC-FDMA systems, the constellation points are assigned in the time-domain and the resulting vectors are transformed into the FD to obtain the transmit vectors~$\bms_\omega$. See, e.g., \cite{WYWDCS2014}, for details on SC-FDMA transmission.

\subsection{Wideband Channel Model}
%


\revision{
In wideband MIMO multicarrier systems, the FD channel matrices~$\bHfreq$, $\omega=0,\ldots,W-1$, are directly related to the time-domain (TD) matrices~$\bHtime\in\complexset^{B\times U}$, where $\ell=0,\ldots,W-1$ are the channel ``taps'' in the TD.
We first introduce the model used to characterize the presence of antenna correlation at the BS side\footnote{In massive MU-MIMO systems, the UEs signals are likely uncorrelated as they are spatially well-separated over potentially large cells or UE scheduling avoids correlated UEs; in contrast, the antennas at the BS are typically confined to a small area, which increases the potential for receive-side correlation.} which occurs in the TD.
Specifically, we use the standard correlation model from \cite{paulraj03} and express the $\ell$th TD channel matrix as follows:
\begin{align}\label{eq:channel_cor_model}
\bHtime  = \bR^{1/2} \bHtimeuncor.
\end{align}
Here, $\bHtimeuncor$ represents an uncorrelated TD channel matrix and
$\bR\in \complexset^{B\times B}$ is a correlation matrix that contains ones on the main diagonal and $\delta\in\reals$ on the off-diagonals.
We allow $\delta$ to be either real positive or negative as long as $\delta^2\leq 1$.
We rewrite the correlation matrix as $\bR= (1-\delta)\bI_B + \delta\bm{1}_B$, where $\bI_B$ and $\bm{1}_B$ is the $B\times B$ identity and all-ones matrix, respectively.
We note that in the absence of receive-side correlation, i.e., $\delta=0$, we have $\bR=\bI_B$ and $\bHtime  =  \bHtimeuncor$.
}

\revision{
In order to take into account the practically-relevant case of imperfect CSI at the BS, we assume that the FD channel matrices $\bHfreq$, $\omega=0,\ldots,W-1$, are obtained from the TD matrices~$\bHtime\in\complexset^{B\times U}$, $\ell=0,\ldots,W-1$, via the discrete Fourier transform~\cite{BGP2002} as follows:\footnote{One could improve the channel estimates $\bHfreq$ by exploiting the fact that only the first $L$ taps of $\bHfreq$ are active in the TD. An analysis of such channel estimation algorithms is left for future work.}
%
%
%
\begin{align} \label{eq:channel_imperfect_CSI}
 \bHfreq = \sum_{\ell=0}^{L-1} \bHtime \exp\!\left(\!-\frac{j2\pi\omega \ell}{W}\right) + \sigma \bE_\omega,
\end{align}
where the matrix $\bE_\omega\in\complexset^{B\times U}$ models channel estimation error on subcarrier $\omega$ and the parameter $\sigma\in\reals^+$ determines the intensity of channel-estimation errors; $\sigma=0$ corresponds to the case for perfect CSI.  
We assume that the entries of the matrix~$\bE_\omega$ are i.i.d.\ (across entries and subcarriers) circularly-symmetric complex Gaussian with unit variance.
}
Equation \fref{eq:channel_imperfect_CSI} relies on the assumption that at most $L\leq W$ of the first channel taps are non-zero (or dominant) and the remaining ones are zero (or insignificant), i.e., $\bHtime=\bZero_{B\times U}$ for $\ell=L,\ldots,W-1$.
In practical OFDM and SC-FDMA systems, the maximum number of non-zero channel taps should not exceed the cyclic prefix length (assuming perfect synchronization). Hence, we can safely assume that $L\ll W $ in practical scenarios and for most standards, such as IEEE 802.11n \cite{IEEE11n} or 3GPP-LTE \cite{3gpp36.211}.
%
%




\subsection{Linear and Non-linear Data Detection and Precoding}
\label{sec:detandpre}

\revisiontwo{
In the massive MU-MIMO uplink, linear and non-linear data detection methods were shown to achieve near-optimal error-rate performance \cite{LETM2014,JGMS2015conf}. 
For linear minimum mean-square error (MMSE) equalization, one first computes  the Gram matrix $\bGfreq=\bHfreq^H\bHfreq$ and then, computes an estimate of the transmit vector as $\hat{\bf s}_\omega = (\bGfreq+\bI\frac{\No}{\Es})^{-1}\bHfreq^H \bmy_\omega$, where $\No$ and $\Es$ stand for the noise variance and average energy per transmit symbol, respectively. 
For non-linear data detectors, such as the one in~\cite{JGMS2015conf}, one can operate directly on the Gram matrix $\bGfreq$ and the matched filter $\bH^H_\omega \bmy_\omega$ without any performance loss \cite{jeon2017achievable}.
}

For such algorithms, a direct computation of the Gram matrix  $\bG_\omega$ for every subcarrier results in excessively high complexity. In fact, even by exploiting symmetries,  $2BU^2$ real-valued multiplications are required, which is more than $16$\,k multiplications per subcarrier for a system with $128$ BS antennas and $8$ users.
Furthermore, the hardware design for massive MU-MIMO data detection in \cite{WYWDCS2014} confirms this observation and shows that  computing the Gram matrix dominates the overall hardware complexity and power consumption by at least $2\times$. 

In the downlink, ZF of Wiener filter precoding are most commonly used \cite{LETM2014}. For example, ZF precoding computes $\bmx_\omega=\bH^H_\omega\bGfreq^{-1}\bms_\omega$, where $\bmx_\omega$ is the $B$-dimensional transmit signal and $\bms_\omega$ the data vector. If the Gram matrix $\bGfreq$ has been precomputed for equalization in the uplink phase, then it can be re-used for ZF precoding in the downlink to minimize recurrent operations. Hence,  to minimize the overall complexity of equalization and precoding, efficient ways to compute the Gram matrix $\bGfreq$ on all active subcarriers $\omega\in\Omega$ are required. 
%


\section{Interpolation-based Gram Matrix Computation}
\label{sec:grammatrixcomputation}

%
 %
\revision{
We now discuss exact and approximate interpolation-based methods for low-complexity Gram matrix computation. 
We note that the exact Gram matrix interpolation assumes that we have perfect CSI, so throughout this section, we will assume that $\sigma=0$.
In \fref{sec:errorbound}, however, we will relax the perfect-CSI assumption and study the performance of exact and approximate interpolation methods with imperfect CSI.
}

As a result of \fref{eq:channel_imperfect_CSI}, the Gram matrices in the FD are given by
\begin{align}
 \bGfreq = \sum^{L-1}_{\ell=0}\sum^{L-1}_{\ell^\prime=0}
 \bHtime^H
 \bHtimeprime \exp 
\!\left(
\frac{j2\pi\omega (\ell-\ell^\prime)}{W}
\right)
\label{eq:Grammodel}
\end{align}
for $w=0,1,\ldots,W-1$. 
Given the FD channel matrices $\bHfreq$ for all active subcarriers $\omega\in\Omega$, a straightforward  ``brute-force'' approach simply computes $\bGfreq=\bHfreq^H\bHfreq$ for each active subcarrier $\omega\in\Omega$.  
\revision{In order to reduce the complexity of such a brute-force approach, we next discuss exact and approximate Gram-matrix interpolation methods that take advantage of the facts that (i) the channel matrices (and hence, the Gram matrices) are ``smooth'' (correlated) across subcarriers if $L<W$ and (ii) massive MU-MIMO benefits from the well-known channel hardening effect~\cite{LETM2014,Marzetta10}.}

\subsection{Exact Gram-Matrix Interpolation}\label{sec:exact_inter}

\revision{The Gram matrix $\bGfreq$ in \fref{eq:Grammodel} is a Laurent polynomial matrix in the variable $x_\omega=\exp(j 2\pi\omega/W)$; we refer the reader to~\cite{CB2010} for more details on  Laurent polynomial matrices. 
Hence, we can establish the following result for exact Gram-matrix interpolation; a short proof is given in \fref{app:exactinterp}.
\begin{lem} \label{lem:exactinterp}
The Gram matrices $\bGfreq$ in \fref{eq:Grammodel} for all subcarriers $\omega=0,\ldots,W$ are fully determined by $2L-1$ distinct and non-zero Gram-matrix base-points.
\end{lem}
Consequently, one can interpolate all of the Gram matrices {\em exactly} from only $2L-1$ distinct and non-zero Gram-matrix base-points that have been computed explicitly.}

In order to perform exact interpolation, we first define a set of base points $\setP\subset\Omega$ that contains $\bpset\ge2L-1$ distinct subcarrier indices. 
We denote the $k$th base-point index as $p_k$, where $k=0,\ldots,\bpset-1$, and the set of all base-point indices as $\setP=\{p_0,\ldots,p_{\bpset-1}\}$.
For each subcarrier index in the base-point set $\setP$, we then explicitly compute $\bpset\geq 2L-1$ Gram matrices~$\bGfreq=\bH_\omega^H\bH_\omega$, $\omega\in\setP$, and perform   entry-wise interpolation for the gram matrices $\bGfreq$ on all remaining active  subcarriers $\omega\in \Omega\backslash\setP$.

The exact interpolation procedure for each entry is as follows. 
For a fixed entry $(m,n)$, we define the vector $\bmg_\setP\in\complexset^\bpset$, which is constructed from the  $(m,n)$ entries $[\bGfreq]_{m,n}$ taken from base-points $\omega\in\setP$, i.e., $\bmg_\setP = \big[ [\bG_{p_0}]_{m,n} \, \cdots \, [\bG_{p_{\bpset-1}}]_{m,n}\big]^T$.  
Then, the vector $\bmg_{\Omega\backslash\setP}\in\complexset^{\abs{\Omega}-\bpset}$ that contains the entry $(m,n)$ for all remaining Gram matrices $\bGfreq$, $\omega\in\Omega\backslash\setP$, is given by
\begin{align}\label{eq:gram_interp}
\bmg_{\Omega\backslash\setP} = \bF_{{\Omega\backslash\setP},L} (\bF_{\setP,L}^\dag \bmg_\setP).
\end{align}
Here, $\bF_{\setP,L}$ represents a $\bpset\times (2L-1)$ matrix where we take the $\bpset$ rows indexed by $\setP$ and the first $L$ and last $L-1$ columns from the $W$-point discrete Fourier transform (DFT) matrix; the entries of the DFT matrix are defined as 
$[\bF]_{m,n} = \frac{1}{\sqrt{W}}\exp\!\left(-\frac{j2\pi}{W}(m-1)(n-1)\right)$.
Similarly, $\bF_{\Omega\backslash\setP,L}$ represents a $(\abs{\Omega}-\bpset)\times (2L-1)$ matrix where we take $\abs{\Omega}-\bpset$ rows indexed by $\Omega\backslash\setP$ and 
the first $L$ and last $L-1$ columns from the $W$-point DFT matrix.
In words, the exact interpolation method in \fref{eq:gram_interp} first computes the $2L-1$ TD Gram-matrix entries  and then, transforms these elements into the frequency domain via the DFT. \revision{See \cite{CBBHB2005,BB2004,CB2011,CB2010} for additional details on other exact interpolation methods developed for MIMO systems.}

Although the method in \fref{eq:gram_interp} is able to \emph{exactly} interpolate the Gram matrix across all $W$ subcarriers, it is in many situations not practical due to the high complexity of the matrix inversion required in $\bF_{\setP,L}^\dag \bmg_\setP$. 
If, however, one can sample the base points uniformly over all $W$ tones, the complexity of matrix inversion can be reduced significantly. Unfortunately, this approach is often infeasible in practice due to the presence of guard-band constraints in OFDM-based or SC-FDMA-based standards \cite{IEEE11n,3gpp36.211}. 
Another issue of exact interpolation methods, such as the ones in \cite{CBBHB2005,BB2004,CB2011,CB2010} and ours in~\fref{eq:gram_interp}, is that they generally assume perfect CSI \revision{and no BS-antenna correlation}. 
As we will show in \fref{sec:error_rate_perf}, imperfect CSI results in poor interpolation performance---this is due to the fact that the matrix $\bF_{\setP,L}$ is \revision{typically} ill-conditioned, especially when sampling Gram-matrices close to the minimum number of $2L-1$ base points.

We next propose two approximate interpolation schemes that not only require (often significantly) lower complexity than a brute-force approach or exact interpolation in~\fref{eq:gram_interp}, but also approach the performance of a brute-force approach in massive MU-MIMO systems and are robust to channel-estimation errors.



\begin{figure}[t]
\centering
\includegraphics[width=0.75\columnwidth]{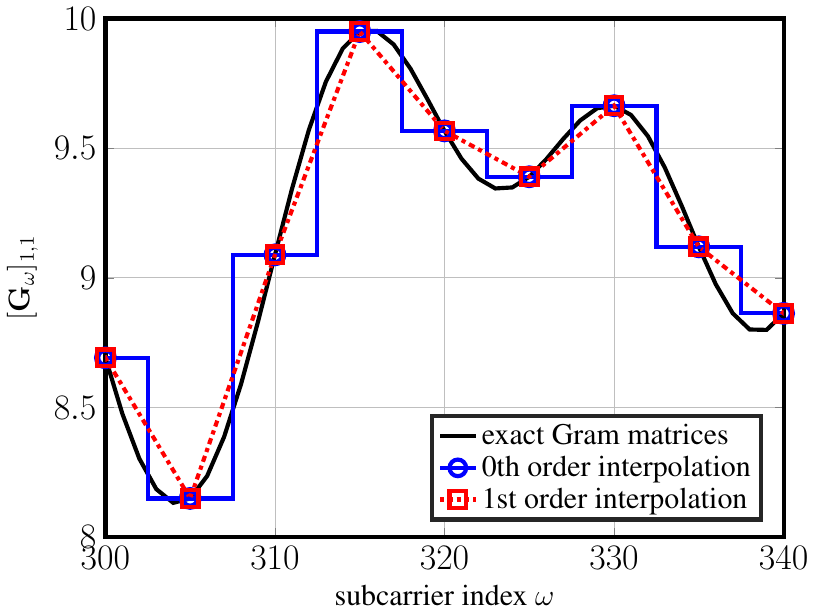}
\caption{Illustration of  \zeroth and \first order interpolation for the entry $[\bG_\omega]_{1,1}$ across subcarriers. We explicitly compute the Gram matrix for every fifth subcarrier index $\omega$ and interpolate the remaining matrices. We assume a 128 BS antenna, 8 user massive MU-MIMO system with $W=2048$ subcarriers, a delay spread of $L=144$, \revision{and perfect CSI without BS-antenna correlation}. 
}
\label{fig:interp_illus}
\end{figure}

\subsection{Approximate Gram-Matrix Interpolation} 
\label{sec:approximateinterpolation}
We consider the following two approximate Gram-matrix interpolation methods illustrated in \fref{fig:interp_illus}.
\subsubsection*{\zeroth Order Interpolation} 
We select a set of $\bpset$ distinct base-points with $\setP=\{p_0,\ldots,p_{\bpset-1}\}\subset\Omega$. 
We explicitly compute $\bG_{p}=\bH^H_{p}\bH_{p}$ on these base points and perform \zeroth order (or nearest-neighbor) interpolation for the remaining subcarriers in the set $\Omega\backslash\setP$ according to:
\begin{align}
\label{eq:0orderinterp}
\bGfreqinterp = \bG_{p}, \quad p = \argmin_{\tilde p\in\setP}| \tilde p-\omega|, \quad \forall \omega\in\Omega\backslash\setP.
\end{align}
In words, we set the interpolated Gram matrix $\bGfreqinterp$ equal to the nearest Gram matrix that has been computed explicitly for one of the neighboring base points.

\subsubsection*{\first Order Interpolation} 
Analogously to the \zeroth order interpolation method, we explicitly compute  $\bG_{p}=\bH^H_{p}\bH_{p}$ on a selected set of base-points $p\in\setP$. 
Then, for each target subcarrier $\omega\in\Omega\backslash\setP$ we pick two nearest base-points $p_k$ and $p_{k+1}$, i.e., $p_k\leq \omega\leq p_{k+1}$, and perform entry-wise linear interpolation according to 
\begin{align}
\bGfreqinterp = \lambda_\omega\bG_{p_k}+(1-\lambda_\omega)\bG_{p_{k+1}}, \quad \omega\in\Omega\backslash\setP,
\label{eq:1orderinterp}
\end{align}
where $\lambda_\omega = (p_{k+1}-\omega)/(p_{k+1}-p_k)$ and $p_{k} < p_{k+1}$.

%

\section{Approximation Error Analysis}
\label{sec:errorbound}

We now analyze the approximation error associated with the approximate  interpolation schemes from \fref{sec:approximateinterpolation}.
We use~$\bGfreq$ to represent the Gram matrices that have been computed exactly and  $\bGfreqinterp$ to represent the Gram matrices that are obtained via approximate interpolation.
Evidently, the exact interpolation scheme in \fref{sec:exact_inter} entails no approximation error.

\subsection{Mean-Square-Error of Approximate Interpolation}
We study the mean-squared error (MSE) on each entry~$(m,n)$ for the $\omega$-th subcarrier, which we define as follows:
\begin{align}
o\textsf{-MSE}^{(m,n)}_\omega\triangleq 
\mathbb{E}
\!\left[
\big|[\bGfreqinterp]_{m,n}-[\bG_\omega]_{m,n}\big|^2
\right]\!.
\label{eq:errordef}
\end{align}
Here, $o$ represents the order of interpolation, i.e., we have either $o=0$ or $o=1$.
Our results make extensive use of the scaled Fej\'er kernel \cite{BN1971} given by 
%
\begin{align}
f_L(\phi) = L^{-2}\frac{1-\cos(L\phi)}{1-\cos(\phi)} 
= L^{-2}
\frac{
\sin^2
(L\phi/2)
}{
\sin^2 
(\phi/2)
}
\label{eq:fejer}
\end{align}
and rely on the following key properties of this kernel; the proof is given in \fref{app:Fejer1}.
%
\begin{lem}\label{lem:Fejer1}
The scaled Fej\'er kernel  \fref{eq:fejer} is non-negative, bounded from above by~one, and monotonically decreasing in~$\phi$ for $\phi\in[0,2\pi/L]$ with $L>1$.
\end{lem}
%

\subsection{MSE of \zeroth Order Interpolation} \label{sec:zerothorder}

The following result precisely characterizes the MSE of \zeroth order interpolation for 
\revision{
imperfect CSI as in \fref{eq:channel_imperfect_CSI} and BS-antenna correlation as in \fref{eq:channel_cor_model};
} 
the proof is given in \fref{app:0ordererror}.
\revision{
\begin{thm} \label{thm:0ordererror}
%
Let the entries of the TD matrices $\bHtimeuncor$, $\ell=1,\ldots,L$, be distributed $\setC\setN(0,1/(BL))$ per complex entry.
Assume that the off-diagonal of the receive correlation matrix be $\delta$,
the variance of the channel estimation error to be $\sigma$, and  
%
$p\in \setP$ is the closest base point to the target subcarrier~$\omega$. 
Then, for any $(m,n)$ entry of the Gram matrix $\bGfreq$, the MSE for the \zeroth order interpolation method in \fref{eq:0orderinterp} is given by
\begin{align}
\MSEzero_\omega
&= \varepsilon_\text{CSI} +\frac{2}{B}(1+ \varepsilon_\text{cor})
\!\left(
1-f_L
\bigg( \frac{2\pi}{W}(p-\omega)
\bigg)
\right)\!,
\label{eq:0ordererrorb}
\end{align}
where we use the definitions
\begin{align*}
 \varepsilon_\text{CSI} = 2 \sigma^2(2+B\sigma^2)\quad \text{and} \quad \varepsilon_\text{cor} = \delta^2(B-1).
\end{align*}
\end{thm}
}

\revision{
From this result, we observe that, as the number of BS antennas~$B$ increases, $\varepsilon_\text{CSI}$ increases quadratically with respect to $\sigma^2$. 
For perfect CSI, i.e., $\sigma=0$ so $\varepsilon_\text{CSI}=0$, the MSE for \zeroth order interpolation decreases with an increasing number of BS antennas~$B$ 
as
$\frac{\partial}{\partial B} \MSEzero_\omega < 0$, if $\delta^2<1$.
} 
\revision{
  Also, we note in the case for non-zero correlation, i.e., $\delta\neq0$, the MSE for \zeroth order interpolation is amplified (compared to that with no correlation) by a factor of $1+\delta^2(B-1)$.
}
Furthermore, we observe that the MSE is independent of the entry of the Gram matrix (i.e., the MSE is identical for the diagonal as well as off-diagonal entries); this is a consequence of the i.i.d.~assumption of the TD channel matrices \revision{$\bHtimeuncor$}.

To gain additional insight into the behavior of $0$th order interpolation in the large BS-antenna limit, we have the following result.
\revision{
\begin{corollary}\label{cor:0_limit}
Assume the conditions in \fref{thm:0ordererror}, and let \revisiontwo{$\sigma^2\to 0$}.
%
Then, as $B\to\infty$, the MSE of \zeroth order interpolation is given by
\begin{align*}
\lim_{B\to\infty}\MSEzero_\omega = 2\delta^2
\!\left(
1-f_L\!
\left( \frac{2\pi}{W}(p-\omega)
\right)\!
\right)\!.
\end{align*}
\end{corollary}
\fref{cor:0_limit} demonstrates that in the large-BS antenna limit, the MSE of \zeroth order interpolation is zero across all subcarriers if and only if the BS antennas are uncorrelated, i.e., $\delta=0$.
For $\delta\neq0$, the MSE depends on the distance between the nearest base-point and the target subcarrier.
}

\revision{While the MSE expression in \fref{eq:0ordererrorb} is exact, it does not provide much intuition.} 
We define the following quantity that enables us to further analyze the MSE in \fref{eq:0ordererrorb}.
\begin{defi} The maximum distance between any subcarrier~$\omega$ and the nearest base point is given by: 
\begin{align}\label{eq:dmax}
\dmax = \max_{k\in\setP}
\left\lfloor
{\frac{p_{k+1}-p_k}{2}} 
\right\rfloor\!.
\end{align}
\end{defi}
With the maximum distance  $\dmax$ for a given set of base points $\setP$, \fref{cor:0th_order_bound} shows that the \zeroth order approximation can be bounded from above using simple analytic expressions; the proof is given in \fref{app:0orderbound}.

\begin{corollary}\label{cor:0th_order_bound}
\revision{Let $\dmax$ be the maximum distance in \fref{eq:dmax} and assume the conditions in \fref{thm:0ordererror}  hold. 
Then, the maximum MSE of \zeroth order interpolation over all active subcarriers $\omega\in\Omega$ is bounded by:}
\begin{align*}
&\max_{\omega\in\Omega}\{\MSEzero_\omega\} 
\\
&\leq 
\begin{dcases}
\varepsilon_\text{CSI} + 
\frac{2}{B}(1+
\varepsilon_\text{cor}), & \dmax \geq \frac {W}{L} \\
\varepsilon_\text{CSI} + 
\frac{2(1+\varepsilon_\text{cor})
}{B} \!\left(
1- f_L\!\left(\frac{2\pi}{W}\dmax\right)\!\right)\!, & \dmax < \frac {W}{L}.
\end{dcases}
\end{align*}
\end{corollary}

\fref{cor:0th_order_bound} implies that regardless of small or large maximum distance $\dmax$, the MSE given by the \zeroth order approximation always decreases with the number of BS antennas~$B$ \revision{if $\delta=0$ and $\sigma^2=0$} (see also \fref{thm:0ordererror}). 
In addition, if the distance between the interpolated subcarrier index $\omega$ and its closest base point is sufficiently small, i.e., $\dmax <  {W}/{L}$, then we obtain a sharper upper bound on the MSE than \revision{$
\varepsilon_\text{CSI} + 
\frac{2}{B}(1+
\varepsilon_\text{cor})$}.
In a scenario with a large delay spread $L$, \fref{cor:0th_order_bound} reveals that one requires finer-spaced base points for \zeroth order interpolation in order to keep the approximation error strictly smaller than \revision{$
\varepsilon_\text{CSI} + 
\frac{2}{B}(1+
\varepsilon_\text{cor})$}. 
Since the maximum error is mainly determined by~$\dmax$, a good strategy for selecting base points with \zeroth order approximation  is uniformly spacing them in the set of active subcarriers $\Omega$.



\subsection{MSE of \first Order Interpolation}

We now present the approximation error analysis of \first order interpolation.
The following result characterizes the MSE of \first order interpolation; the proof is given in \fref{app:1ordererror}.
\revision{
\begin{thm} \label{thm:1ordererror}
Let the entries of the TD matrices $\bHtimeuncor$, $\ell=1,\ldots,L$, be distributed $\setC\setN(0,1/(BL))$ per complex entry.
Assume that the off-diagonal of the receive correlation matrix be $\delta$,
the variance of the channel estimation error to be $\sigma$ across all subcarriers $\omega=0,\ldots,$, and  
%
$p\in \setP$ is the closest base point to the target subcarrier $\omega$. 
Then, for any $(m,n)$-th entry of the Gram matrix $\bGfreq$, the MSE for the \first order interpolation method in \fref{eq:0orderinterp} is given by
\begin{align}
&\MSEone_\omega  = \varepsilon_\text{CSI}( 1 - \lambda_\omega(1- \lambda_\omega ))
\notag\\&   +  \frac{2}{B}
(1+\varepsilon_\text{cor})
\big( 1 -\lambda_\omega(1- \lambda_\omega) + \lambda_\omega(1-\lambda_\omega) f_L(\theta) \notag \\
& -(1-\lambda_\omega)f_L(\lambda_\omega\theta)-\lambda_\omega f_L((1-\lambda_\omega)\theta)\big),
\label{eq:1ordererrorb}
\end{align}
where $\theta=\frac{2\pi}{W}(p_{k+1}-p_k)$ and $\lambda_\omega= (p_{k+1}-\omega)/(p_{k+1}-p_k)$.
\end{thm}
}

\revision{
Analogously to \zeroth order interpolation, we observe that the MSE of \first order interpolation is independent of the entry  $(m,n)$ and impacted by CSI errors and receive correlation (see \fref{sec:zerothorder} for detailed discussion).
}
The result shown next in \fref{cor:1orderbound} reveals that if the spacing between the two base-points $p_k$ and $p_{k+1}$ defined as 
$d_k=p_{k+1}-p_k$ is sufficiently small, then the \first order interpolation strictly outperforms \zeroth order interpolation, i.e., $\MSEone_\omega<\MSEzero_\omega$ for all $\omega\in(p_k,p_{k+1})$; the proof is given in \fref{app:1orderbound}.

\begin{corollary} 
Let $d_k$ denote the spacing between two base-points $p_k$ and $p_{k+1}$, and  assume the conditions in \fref{thm:0ordererror} and  \fref{thm:1ordererror}  hold. If $d_k \leq W/(3L)$, then 
\begin{align}
\MSEone_\omega \leq \MSEzero_\omega, \quad \text{ for all } \omega\in(p_k,p_{k+1}),
\end{align}
which holds with equality if and only if $L=1$ \revision{and $\varepsilon_\text{CSI}=0$.}
\label{cor:1orderbound}
\end{corollary}

We note that the condition $d_k<W/(3L)$ is not sharp; \fref{app:1orderbound} outlines the details on how it can be sharpened.
Furthermore,  given that $d_k$ is significantly larger than $W/(3L)$, we can construct situations for which \zeroth order interpolation \emph{outperforms} \first order interpolation. Note that for $L=1$, the FD channel is flat (i.e., $\bG_\omega$ is constant for all $\omega$) and hence, \first and \zeroth order interpolation have the same MSE.

In summary, we observe that for both approximate interpolation methods, the MSE can be lowered by increasing the number of BS antennas $B$ \revision{assuming that the channel estimation error~$\varepsilon_\text{CSI}$ decreases with $B$.} 
\revision{In the large-antenna limit $B\to\infty$ with perfect CSI and no BS-antenna correlation, the MSE vanishes, which is an immediate consequence of channel hardening in massive MU-MIMO systems~\cite{LETM2014}.}
Furthermore, \first order interpolation generally outperforms \zeroth order interpolation for a sufficiently small minimum spacing between adjacent base points, i.e., for $d_k\leq W/(3L)$.




\section{Complexity Analysis}\label{sec:complexityanalysis}

We next compare the computational complexity of the four studied Gram-matrix computation algorithms: brute-force computation, exact interpolation, \zeroth order interpolation, and \first order interpolation.
We measure the computational complexity by counting the number of real-valued multiplications.\footnote{We assume that a complex-valued multiplication requires four real-valued multiplications; computation of the squared magnitude of a complex number is assumed to require two real-valued multiplications.}
%
%
%

\subsection{Brute-Force Computation}

We start by deriving the total computational complexity required by the brute-force (BF)  method.
We only compute  the upper triangular part of $\bGfreq$ (since the matrix is Hermitian).
Each off-diagonal entry requires $B$ complex-valued multiplications, which corresponds to $4B$ real-valued multiplications; each diagonal entry requires only $2B$ real-valued multiplications. 
Hence, the computational complexity of computing $\bGfreq$ using the BF method is  
\begin{align}
C_{\textnormal{BF}} & = \activeset\!\left(4B \frac{U(U-1)}2+2B U\right)\! = 2\activeset BU^2
\label{eq:bfcost}
\end{align}
for  a total number of $\activeset $ active subcarriers. 

\subsection{Exact Interpolation}\label{sec:subsec_exact}
We now derive the computational complexity of exact interpolation as discussed in \fref{sec:exact_inter}.
%
Exact interpolation requires a BF computation of the Gram matrix at each of the $\bpset$ base points.
We will use the $\bpset$ precomputed base points of $\bGfreq$ to interpolate the  remaining $\activeset -\bpset$ Gram matrices.

We will assume that the base points and the assumed channel delay spread $L$ are fixed \emph{a-priori} so that~$\bF_{\Omega\backslash\setP,L}\bF_{\setP,L}^\dag$ in~\fref{eq:gram_interp} can be precomputed and stored. 
We emphasize that this approach does not include the computational complexity of computing the interpolation matrix itself, which favors this particular interpolation scheme from a complexity perspective. 
In fact, we only need to multiply the precomputed interpolation matrix $\bF_{\Omega\backslash\setP,L}\bF_{\setP,L}^\dag$ with the vector $\bmg_\setP$, which requires $4(\activeset -\bpset)\bpset$ real-valued multiplications. 
Hence, the total computational complexity of exact interpolation is:
\begin{align}
C_{\textnormal{Exact}} & = \frac{\bpset}{\activeset }C_{\textnormal{BF}}
+ 4(\activeset -\bpset)\bpset\frac{U(U+1)}2 \notag \\
& = 2\bpset(\activeset -\bpset+B)U^2  +2\bpset(\activeset -\bpset)U.
\label{eq:exactcost}
\end{align}
We note that if the number of users $U$ is large and the number of base points is similar to the number of BS antennas, i.e., $\bpset\simeq B$, then the BF method in \fref{eq:bfcost} and exact  interpolation~\fref{eq:exactcost} exhibit similar complexity.
We also observe that the complexity of exact interpolation \fref{eq:exactcost} is lower than that  of the BF method~\fref{eq:bfcost} if 
$\bpset<(1+U)^{-1}BU$.
Since the use of  $\bpset\geq 2L-1$ distinct base points guarantees exact interpolation (assuming perfect CSI), we observe that exact interpolation has lower complexity than the BF method if $L$ is (approximately) smaller than $B/2$.

%
%
%
%
\subsection{0th Order Interpolation}
%
%
The computational complexity of the \zeroth order interpolation method is given by 
\begin{align}
C_{\textnormal{\zeroth}} = 2\bpset BU^2,
\label{eq:0ordercost}
\end{align}
as we only need to compute the Gram matrices on all the base points. 
\revision{We note that since typically $\bpset \ll \activeset$ the savings (in terms of real-valued multiplications) are significant compared to the BF approach and exact interpolation, but does so at the cost of approximation errors (cf.~\fref{sec:complex_pref_toff}).}


\subsection{1st Order Interpolation}
The computational complexity of the \first order interpolation is given by
\begin{align}
C_{\textnormal{\first}} & = C_{\textnormal{\zeroth}} + 4(\activeset-\bpset)\frac{U(U+1)}2 \notag \\
& = 2\bpset BU^2+2(\activeset -\bpset)U(U+1),
\label{eq:1ordercost}
\end{align}
where we assume that the interpolation weight $\lambda_\omega$ was precomputed.
We note that the linear interpolation stage for each subcarrier $\omega\in\Omega\backslash\setP$ requires four real-valued multiplications.
%
By comparing \fref{eq:0ordercost} to \fref{eq:1ordercost}, we observe that the complexity of \first order interpolation always exceeds the complexity of the \zeroth order method, but the complexity is significantly lower than that of the BF method as we generally have  $\bpset \ll \activeset$.

\section{Numerical Results}\label{sec:numericalresults}
We now study the MSE, the error-rate performance, and the computational complexity of the proposed Gram-matrix interpolation schemes. 
We consider a MU-MIMO-OFDM system with 128 BS antennas and with 8 single-antenna users. We assume a total of $W=2048$ subcarriers, with $|\Omega|=1200$ active subcarriers, similar to that used in 3GPP LTE~\cite{3gpp36.211}.
Unless stated otherwise, we assume that the entries of the TD channel matrices are i.i.d.\ circularly-symmetric complex Gaussian with variance $1/(BL)$ and we consider 16-QAM  transmission (with Gray mapping). 
We use a linear MMSE equalizer for data detection; see \fref{sec:detandpre}.
For situations with imperfect CSI, we consider pilot-based maximum-likelihood (ML)  channel estimation with a single orthogonal pilot sequence of length~$U$ with the same transmit power as for the data symbols.
\revision{
\subsection{Complexity Comparison}
We now assess the complexity of the various Gram-matrix computation methods in comparison to the overall complexity required for linear MMSE-based data detection, which includes Gram-matrix and matched-filter computation as well as matrix inversion for each active subcarrier. 
The results shown here are for a $128\times8$ (the notation represents $B\times U$) massive MU-OFDM-MIMO system with $1200$ active subcarriers and a delay spread of $L=144$.}
%
%
%

\begin{figure}[tp]
\centering
\includegraphics[width=0.75\columnwidth]{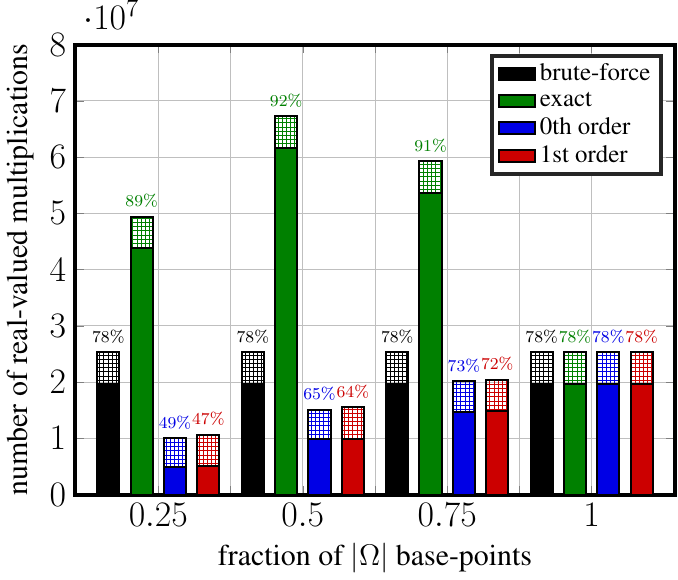}
\caption{
\revision{
  We compare the complexity of various  Gram matrix computation methods in comparison to the complexity of data detection for a $128\times8$ MU-OFDM-MIMO system with $\abs{\Omega}=1200$, $L=144$, and four different sets of base-points $\abs{\setP} = \{0.25\abs{\Omega},0.5\abs{\Omega},0.75\abs{\Omega},\abs{\Omega}\}$.
  The percentage values shown in the above bar plots show the relative percentage of Gram-matrix computation compared to the total  complexity required for data detection.
}
%
%
  }
\label{fig:gram_graph}
\end{figure}

\revision{
\fref{fig:gram_graph} compares the complexity of Gram matrix computation for four different methods, brute-force, exact, 0th-, and 1st-order interpolation methods for $\abs{\setP} = \{0.25\abs{\Omega},0.5\abs{\Omega},0.75\abs{\Omega},\abs{\Omega}\}$. 
The solid part of the bar plot shows the complexity of Gram matrix computation; the fenced part corresponds to the remaining complexity required for data detection (including matched-filter computation and a matrix inversion for each active subcarrier). The percentage values indicate the relative complexity of Gram-matrix computation compared to the total complexity required for data detection.
\revisiontwo{
We assume a Cholesky-based implicit matrix inversion for detection \cite{WYLDCS2018}. 
As demonstrated in \cite{studer2011asic},  Gram matrix computation requires majority of the computational complexity, as it scales quadratically in the number of BS antennas.
}%
}

\begin{figure*}[tp]
\centering
\subfigure[Perfect CSI (i.i.d.\ Rayleigh fading).]{
\includegraphics[width=0.62\columnwidth]{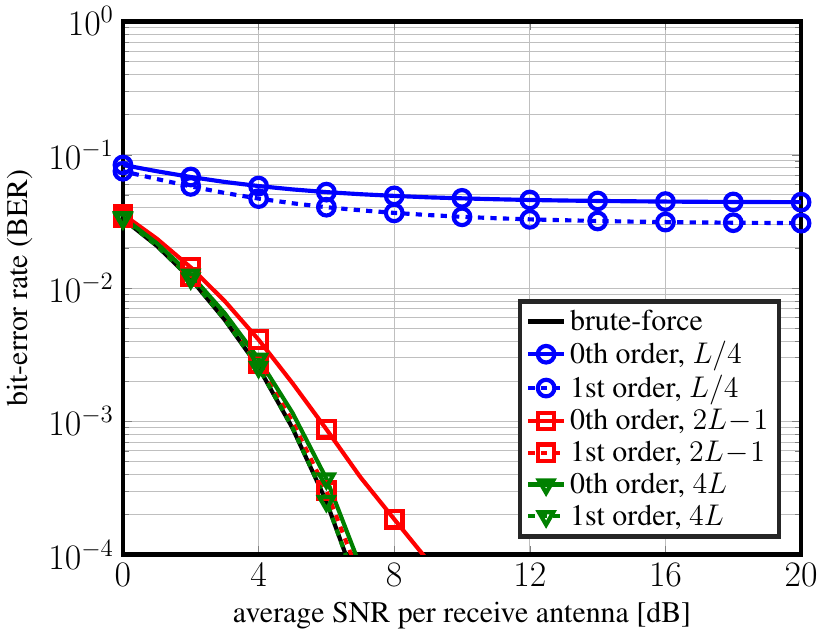}
\label{fig:bitA}}
\hspace{0.1cm}
\subfigure[Imperfect CSI (i.i.d.\ Rayleigh fading).]{
\includegraphics[width=0.62\columnwidth]{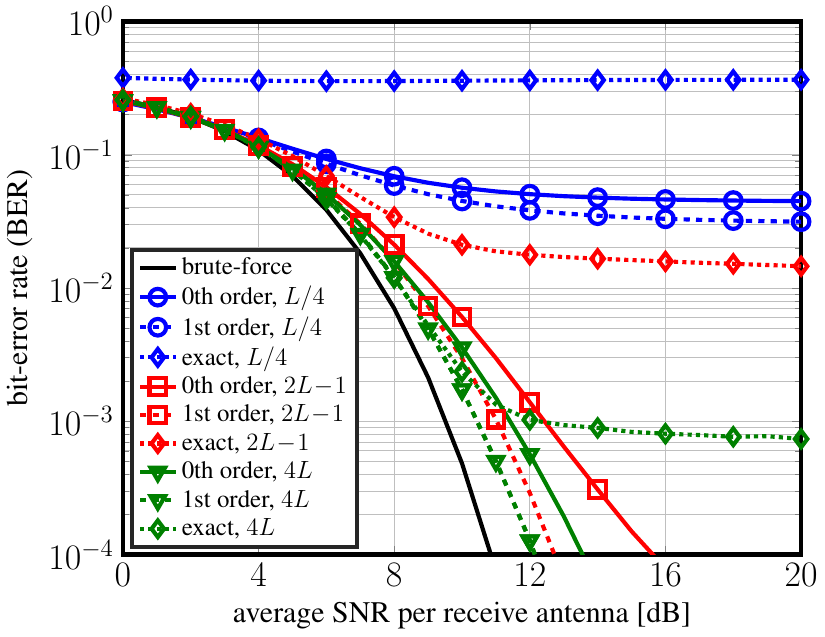}
\label{fig:bitB}}
\hspace{0.1cm}
\subfigure[Imperfect CSI (QuaDRiGa channel model).]{
\includegraphics[width=0.62\columnwidth]{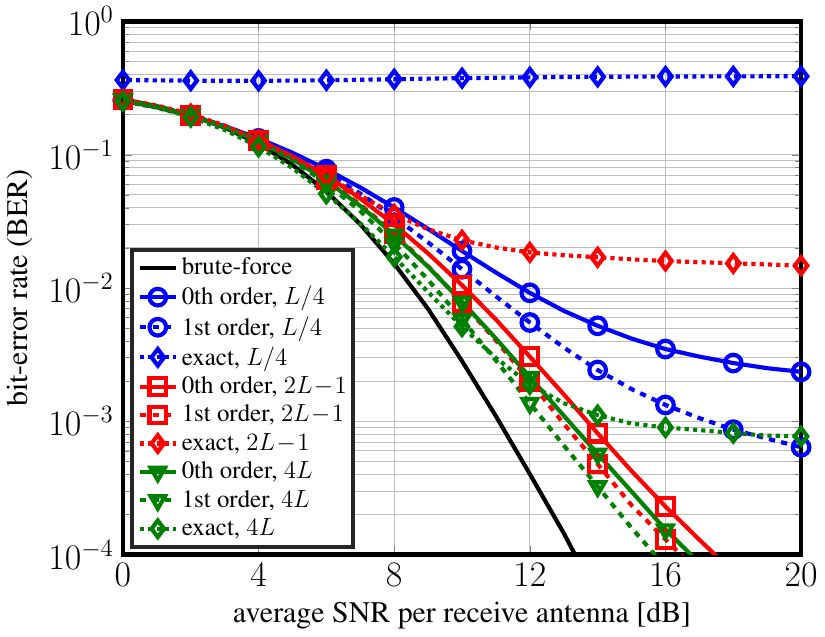}
\label{fig:bitC}}
\caption{
Uncoded bit error-rate (BER) comparison in a 128 BS antenna, 8 (single-antenna) user, wideband massive MU-MIMO-OFDM system. The values next to the legend entries correspond to the number of base points $\abs{\setP}$.
The proposed approximate interpolation schemes (\zeroth and \first order interpolation) outperform exact interpolation for scenarios with imperfect CSI and approach the performance of the exact brute-force method for a small number of base points.}
\label{fig:biterrorrate}
\end{figure*}

\revision{We see that the exact interpolation method results in high complexity in the considered system (see \fref{sec:subsec_exact} for exact details when exact interpolation achieves lower complexity than a BF approach).
We also see that the proposed 0th and 1st order approximation methods both achieve significant complexity reductions. For $\abs{\setP} = 0.25\abs{\Omega}$, the proposed methods requires less than half the complexity of a BF approach. 
As we will show in \fref{sec:error_rate_perf}, the proposed approximate interpolation methods will exhibit similar error-rate performance as that the BF approach (see Figs. \ref{fig:biterrorrate} and \ref{fig:tradeoff}), but does so at fraction of the computational complexity.
%
%
%
%
}
%

\subsection{MSE of Approximate Interpolation}
\begin{figure}[t]
\centering
\includegraphics[width=0.75\columnwidth]{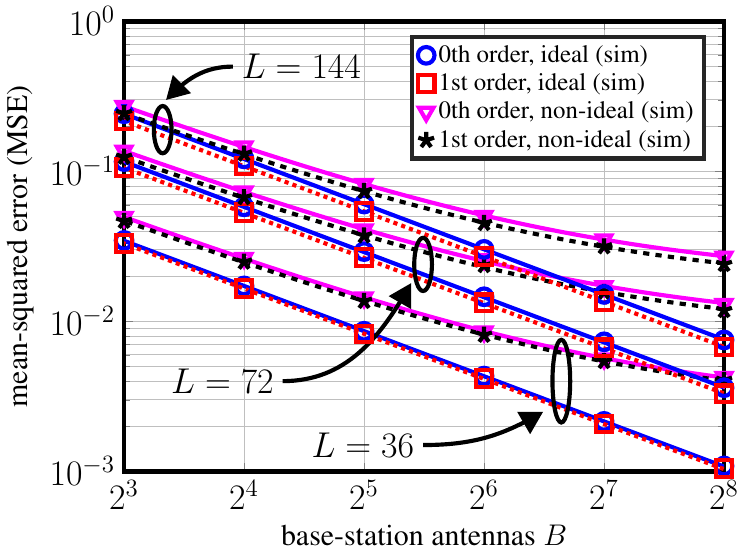}
\caption{
MSE of \zeroth and \first order interpolation for an entry of $\bG_{512}$ using two base points $\bG_{500}$ and $\bG_{600}$ for three different delay spreads $L\in\{36,72,144\}$ 
\revision{for the ideal and non-ideal scenarios.
The markers represent simulation results whereas the lines represent our approximation-error analysis. Evidently, our theory matches perfectly with the simulated values. 
}
%
}
\label{fig:graminterpolationerror}
\end{figure}

\fref{fig:graminterpolationerror} compares the MSE of \zeroth and \first order interpolation as proposed in \fref{sec:approximateinterpolation}. 
Note that the BF method and exact interpolation have an MSE of zero and hence, we exclude these results.
We select two base points at subcarriers $500$ and $600$ and one target point at subcarrier $512$, and compare the MSE for different numbers of BS antennas \revision{and under ideal and non-ideal scenarios. In the ideal scenario, we assume perfect CSI and no BS-antenna correlation, whereas in the non-ideal scenario we assume channel-estimation at $\SNR=25$~dB across all subcarriers with the signal-to-noise ratio (SNR) defined by $\SNR = U/(B\sigma^2)$ and a BS-antenna correlation of $\delta=0.1$.}
In order to assess the approximation error with respect to different channel delay spreads, we set $L\in\{36,72,144\}$. 
The resulting MSE is shown in \fref{fig:graminterpolationerror}.
\revision{Note that the MSE for both \zeroth and \first are independent of the entry (as predicted by Theorems~\ref{thm:0ordererror} and \ref{thm:1ordererror}); hence, we consider the average MSE across all entries.}

We observe that the \first order interpolation method achieves a lower MSE than that given by \zeroth order interpolation, where the performance gap increases with larger delay spreads $L$. 
This observation is caused by the fact that for small delay spreads $L$, the channel is more smooth across subcarriers.
For larger delay spreads $L$, \first order interpolation captures the faster-changing behavior of the Gram matrix, whereas the \zeroth order interpolation ignores such changes.
\revision{We also see that the MSE degrades in the non-ideal scenario, even if we increase the number of BS antennas; this behavior is reflected in our  analytical results.}
Finally, we see that the simulated MSE matches perfectly our theoretical results in Theorems \ref{thm:0ordererror} and \ref{thm:1ordererror}.

\subsection{Error-rate Performance}\label{sec:error_rate_perf}
We now compare the error-rate performance of the proposed Gram-matrix computation schemes. 
We simulate the bit-error rate (BER) for a MU-MIMO-OFDM system for a different number of base-points~$\bpset$ and for perfect as well as imperfect CSI. We also investigate the impact of a more realistic channel model. 
For all results, we simulate three different numbers of base-points $\bpset=L/4$, $\bpset=2L-1$, and $\bpset=4L$, and select equally-spaced base points.
Figures~\ref{fig:bitA} and \ref{fig:bitB} show BER simulation results for an i.i.d.\ Rayleigh fading scenario with perfect and imperfect CSI, respectively. Figure~\ref{fig:bitC} shows BER simulation results for the QuaDRiGa channel model\footnote{We simulate a square antenna array with a non-line-of-sight scenario with a $2$\,GHz carrier frequency, $20$\,MHz bandwidth, and $200$\,m distance between BS antenna and the users. Our algorithms assume $L=144$ but the true delay spread is slightly smaller.} with imperfect CSI~\cite{JRBT2014}.
\revisiontwo{We note that QuaDRiGa channel model includes a path-loss model for each user.}

Figure~\ref{fig:bitA} shows that exact interpolation for $\bpset\geq 2L-1$ base points provides identical results as the BF method (up to machine precision) for a system with perfect CSI. 
For $\bpset=L/4$ base points, the proposed \zeroth and \first order interpolation exhibit an error floor; this performance loss can be  mitigated substantially by increasing the number of base points to $\bpset=2L-1$. 
By setting $\bpset=4L < 0.5 \activeset $, the both the \zeroth and \first order interpolation methods exhibit virtually no BER performance loss.

Figure~\ref{fig:bitB} shows  the situation for imperfect CSI (with channel estimation). We observe that the performance of the BF method and that of exact interpolation are no longer equal. 
In fact, for $\bpset=2L-1$ and $\bpset=4L$ base points, exact interpolation exhibits a significant error floor.
The reason is due to the fact that the interpolation matrix is ill-conditioned, which results in  significant noise enhancement artifacts. 
Although the error floor is decreased for $\bpset=4L$, a floor remains at  $10^{-3}$ BER.
In contrast, the error floor of \zeroth order and \first order interpolation for $\bpset=2L-1$ and $\bpset=4L$ base points is well-below $10^{-4}$ BER and hence, the proposed approximate interpolation schemes are more resilient to scenarios with imperfect CSI than exact interpolation. 

Figure~\ref{fig:bitC} shows the BER performance for the QuaDRiGa channel model with imperfect CSI.
We observe that all considered interpolation methods achieve a lower error floor than that given in \fref{fig:bitB} for $\bpset=L/4$; this is due to the fact that the effective delay spread for the considered channel is smaller than $L=144$ (which is assumed in the algorithms). 
Once again, we observe a BER floor of exact interpolation for all considered numbers of base points. 
In summary,   the proposed approximate interpolation methods are more robust in practical scenarios than the exact interpolation method.

\begin{figure}[tp]
\centering
\includegraphics[width=0.75\columnwidth]{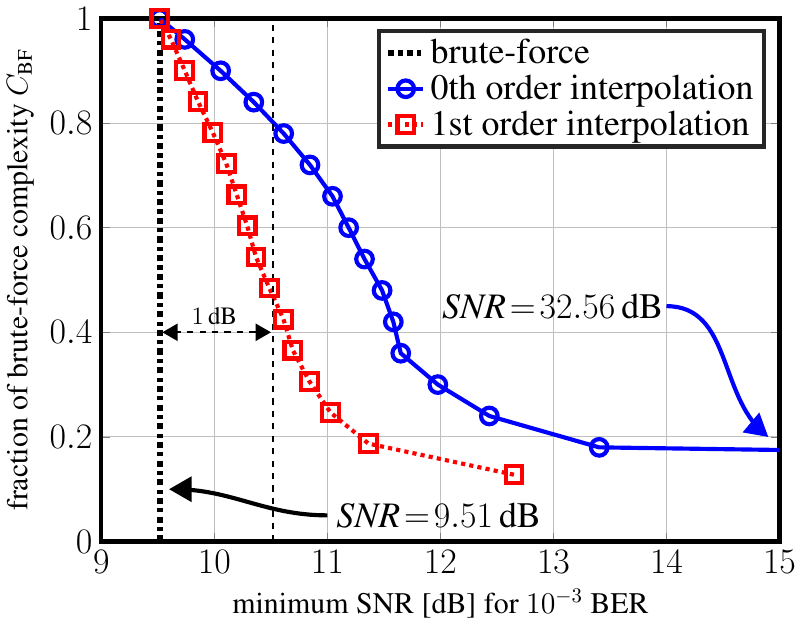}
\caption{Trade-off between SNR performance and computational complexity for \zeroth and \first order interpolation in an i.i.d.\ Rayleigh fading channel with imperfect CSI. 
Both approximate interpolation methods approach the performance of an exact brute-force approach at a fraction of the complexity.}
\label{fig:tradeoff}
\end{figure}

\subsection{Performance/Complexity Trade-off}
\label{sec:complex_pref_toff}

We now investigate the BER performance vs.\ computational complexity trade-off for the proposed approximate interpolation methods with imperfect CSI. 
We use the complexity  $C_\textnormal{BF}$ of the BF method in \fref{eq:bfcost} as our baseline, and we compare it to that of the proposed \zeroth and \first order interpolation methods in~\fref{eq:0ordercost} and \fref{eq:1ordercost}, respectively.
We vary the number of base points $\bpset$ from $L$ to $D$ and simulate the minimum SNR required for the linear MMSE equalizer to achieve $10^{-3}$ BER.

Figure~\ref{fig:tradeoff} shows the trade-off results for \zeroth and \first order interpolation. For a fixed fraction of the complexity of $C_\textnormal{BF}$, we observe that the \first order interpolation method always outperforms the \zeroth order interpolation method.
Hence, \fref{fig:tradeoff} clearly reveals that the additional complexity required by linear interpolation is beneficial when jointly considering performance and complexity.
In addition, we see that the \first order interpolation method approaches the SNR performance of the BF method by 1\,dB with only $45\%$ of the complexity.

\section{Conclusions}
\label{sec:conclusions}

We have studied the performance of exact and approximate interpolation-based Gram matrix computation for wideband massive MU-MIMO-OFDM systems. 
Instead of performing a brute-force (BF) computation of the Gram matrix for all subcarriers or using exact interpolation schemes, we have proposed two simple, yet efficient approximate interpolation methods. 
We have demonstrated that channel hardening in massive MU-MIMO enables the proposed \zeroth and \first order interpolation schemes to perform  close to that of an exact BF computation at only a fraction of the computational complexity.
In addition, the proposed approximate interpolation methods are more robust to \revision{channel-estimation errors and receive-side antenna correlation} than exact interpolation methods. 

\revision{There are many avenues for future work. We expect the use of higher-order approximate Gram-matrix interpolation schemes to perform better at an increase in complexity. 
An analysis of such methods is left for future work. Our results also indicate that the broad range of existing exact interpolation schemes for small-scale, point-to-point MIMO systems, e.g., matrix inversion and QR decomposition~\cite{BB2004,CBBHB2005} can be made more robust and less complex in massive MU-MIMO systems if combined with approximate, low-order interpolation schemes. In fact, the recent result in \cite{MAMMOET} for approximate interpolation of matrix inversion demonstrates this claim in a massive MU-MIMO scenario via simulations. 
%
%
In addition, a performance analysis of the proposed or other approximate interpolation methods in more realistic systems that suffer from frequency, timing offset, and \revisiontwo{a more realistic model for receive-side correlation} is left for future work.
}

\section*{Acknowledgments}

The authors would like to thank M.~Wu and C.~Dick from Xilinx, Inc.\ for insightful discussions on interpolation-based matrix computations.
\revision{The authors would also like to thank the anonymous reviewers for their suggestions.}
\revision{The work of C. Jeon and C. Studer was supported in part by Xilinx Inc., and by the US National Science Foundation (NSF) under grants ECCS-1408006, CCF-1535897,  CAREER CCF-1652065, CNS-1717559, and EECS-1824379.}

 

\begin{appendices}

\section{Proof of \fref{lem:exactinterp}}\label{app:exactinterp}
\revision{Since all of the possible values in the exponent of \fref{eq:Grammodel}, i.e., $\ell-\ell'$, are integers ranging from $-(L-1)$ to $L-1$, the Gram matrix $\bGfreq$ in \fref{eq:Grammodel} is a polynomial with degree no larger than $2L-1$. 
Consequently, the Gram matrices $\bGfreq$ in \fref{eq:Grammodel} for all subcarriers $\omega=0,\ldots,W$ are fully determined from $2L-1$ distinct and non-zero Gram-matrix base-points.}

\section{Proof of \fref{lem:Fejer1}}\label{app:Fejer1}
Evidently, \fref{eq:fejer} is non-negative, i.e., $f_L(\phi)\geq0$.
To show that $f_L(\phi)\leq 1$, we use the fact that the Fej\'er kernel $Lf_L(\phi)=L^{-1} \frac{1-\cos (L\phi)}{1-\cos(\phi)}$ is upper bounded by $L$ \cite[Eq.~1.2.24]{BN1971}.
Now, we show that $f_L(\phi)$ is monotonically decreasing in $\phi\in[0,2\pi/L]$ for $L>1$.
We start by defining an auxiliary function $g(\phi)=\sin(L\phi/2)/\sin(\phi/2)$, so that $f_L(\phi)=g_L^2(\phi)/L^2$.
For $L>1$, $\phi\in(\pi/L,2\pi/L]$, $\sin(L\phi/2)$ and $\sin(\phi/2)$ are monotonically decreasing and increasing respectively, and hence, $g(\phi)$ and $f_L(\phi)$, are monotonically decreasing. 
For $\phi\in[0,\pi/L]$, the derivative of $g(\phi)$ with respect to $\phi$ is given by:
\begin{align*}
\frac{\dif g(\phi)}{\dif\phi} = \frac {\left(L\cos(L\phi/2)\sin(\phi/2) - \cos(\phi/2)\sin(L\phi/2)\right)}{2\sin^2(\phi/2)},
\end{align*}
and we have that
\begin{align*}
\frac{L\cos(L\phi/2)\sin(\phi/2)}{\cos(\phi/2)\sin(L\phi/2)} = \frac{L\tan(\phi/2)}{\tan(L\phi/2)}<1.
\end{align*}
Hence, $g^\prime(\phi)<0$, and therefore, $f_L(\phi)$ is monotonically decreasing in $\phi\in[0,2\pi/L]$.




%
%
%
%
%
%
%
%
%
%
\section{Proof of Theorem \ref{thm:0ordererror}}\label{app:0ordererror}

Suppose we use $\bG_{p_k}$ at base point $p_k$ to approximate $\bG_w$ at the target subcarrier index $\omega$. 
Hence, the MSE in \fref{eq:errordef} is given by the following expression:
\begin{align} 
\MSEzero^{(m,n)}_\omega
& = 
\mathbb{E}
\!\left[
\big|[\widetilde\bG_\omega]_{m,n}-[\bG_\omega]_{m,n}\big|^2
\right]\!.
\label{eq:0ordererrordef}
\end{align}
\revision{
We will obtain an analytical expression for \fref{eq:0ordererrordef} with imperfect CSI and  the BS-antenna correlation model  introduced in \fref{eq:channel_imperfect_CSI} and \fref{eq:channel_cor_model}, respectively.
We start with expressing the channel matrix $\bHfreq$ from \fref{eq:channel_imperfect_CSI} as
\begin{align}
\bHfreq
\stackrel{(a)}{=} 
 \sum_{\ell=0}^{W-1}
\!\left(
\bHtime\mathbb{I}(\ell < L) + \frac{\sigma}{\sqrt{W}} \tilde\bE_\ell 
\right)\!
\exp\!\left(-\frac{j2\pi\omega \ell}{W}\right)\!,
\label{eq:Homega_imperfectCSI}
\end{align}
where $(a)$ follows from noting that the DFT is orthogonal (with normalization constant) with each entries of $\tilde\bE_\ell$ distributed $\setC\setN(0,1)$.
Hence, the TD channel matrix under imperfect CSI is given as $\widehat\bH_\ell ^\sigma = \bHtime \mathbb{I}(\ell < L) + \frac{\sigma}{\sqrt{W}} \tilde\bE_\ell $.
}

\revision{
Now, use the correlation model introduced in~\fref{eq:channel_cor_model}. 
We note that the $B\times B$ BS correlation matrix $\bR = (1-\delta)\bI_B + \delta \bm{1}_B$ can be expressed by $(\alpha \bI_B + \beta \bm{1}_B)^2 = \bR$ where $\alpha = \sqrt{1-\delta}$, and $\beta = \frac{1}{B}(\sqrt{1 + (B-1)\delta}-\alpha)$.
Since the matrix $(\alpha \bI_B + \beta \bm{1}_B)$ is symmetric, we note that the $\bHtime$ is expressed as $\bHtime = (\alpha \bI_B + \beta \bm{1}_B)\bHtimeuncor$ so that
\begin{align}
\HtimeIJ=\alpha[\bHtimeuncor]_{m,n} + \beta\sum_{b=1}^B [\bHtimeuncor]_{b,n}
\label{eq:Htime_cor}
\end{align}
For our derivation of the MSE, we will utilize the following auxiliary function:}
\begin{align}
R_{\ell \ell^\prime}=
\exp
\!\left(
{j\frac{2\pi p_k}{W}(\ell-\ell^\prime)}
\right)
-
\exp
\!\left(
j\frac{2\pi \omega}{W}(\ell-\ell^\prime)
\right)\!.
\label{eq:Rdef}
\end{align}
By substituting $R_{\ell \ell^\prime}$ into $[\widetilde{\bG}_\omega]_{m,n}-[\bG_\omega]_{m,n}$ in \eqref{eq:0ordererrordef}, we obtain the following expression for the \zeroth order MSE:
\revision{
\begin{align}
&\MSEzeroIJomega  = 
\mathbb{E}
\bigg[
\bigg\vert
\!\sum^{W-1}_{\ell,\ell^\prime=0}
\sum^B_{b=1}\,
[\bHtimenoidx_\ell^\sigma]_{b,m}^*[\bHtimenoidx_{\ell^\prime}^\sigma]_{b,n} 
R_{\ell \ell^\prime} 
\bigg\vert^2
\bigg] 
\notag \\
& = 
\sum^{W-1}_{\ell_1,\ell_2=0}
\sum^{W-1}_{\ell_3,\ell_4=0}
\sum^B_{b_1,b_2=1}
\!\bigg(
R_{\ell_1 \ell_2}^* R_{\ell_3 \ell_4}
\notag \\
& \mathbb{E}
\big[
[\bHtimenoidx_{\ell_1}^\sigma]_{b_1,m}
[\bHtimenoidx_{\ell_2}^\sigma]_{b_1,n}^*
[\bHtimenoidx_{\ell_3}^\sigma]_{b_2,m}^*
[\bHtimenoidx_{\ell_4}^\sigma]_{b_2,n}
\big]\bigg)
\notag \\
& \stackrel{(a)}{=} 
\sum^{W-1}_{\substack{\ell_1,\ell_2=0 \\\ell_1\neq\ell_2}}
|{R_{\ell_1 \ell_2}}|^2
\notag
\\& 
\sum^B_{b_1,b_2=1} 
\mathbb{E}
\big[
[\bHtimenoidx_{\ell_1}^\sigma]_{b_2,m}^*
[\bHtimenoidx_{\ell_1}^\sigma]_{b_1,m}
\big]
\mathbb{E}
\big[
[\bHtimenoidx_{\ell_2}^\sigma]_{b_1,n}^*
[\bHtimenoidx_{\ell_2}^\sigma]_{b_2,n}
\big]
\label{eq:0ordererroraproof1}
\end{align}
where $(a)$ follows from $R_{\ell_1\ell_2}=0$ if $\ell_1 = \ell_2$, and independence and the zero-mean assumption on the TD channel for $\ell_1\neq\ell_2$, 
which enforces $\ell_1=\ell_3$ and $\ell_2=\ell_4$ for \fref{eq:0ordererroraproof1}.
By inspection of \fref{eq:0ordererroraproof1}, we observe that $\MSEzeroIJomega$ is independent of $m$ and $n$ and thus, the MSE of the off-diagonal and diagonal entries are equal. 
}
\revision{
%
%
We simplify \fref{eq:0ordererroraproof1} for imperfect CSI and the BS-antenna correlation model.
We first note that
\begin{align*}
&\mathbb{E}
\big[
[\bHtimenoidx_{\ell_1}^\sigma]_{b_2,m}^*
[\bHtimenoidx_{\ell_1}^\sigma]_{b_1,m}
\big] = 
\frac{\sigma^2}{W} \mathbb{I}(b_1=b_2)
\\
&+ \frac{1}{BL}(\alpha^2 \mathbb{I}(b_1=b_2) + 2 \alpha\beta + \beta^2 B)\mathbb{I}(\ell _1< L)  
\\
&=
\frac{\sigma^2}{W} \mathbb{I}(b_1=b_2) + \frac{1}{BL}(\alpha^2 \mathbb{I}(b_1=b_2) + \delta)\mathbb{I}(\ell _1< L),
\end{align*}
where the last step is obtained by noting that $2 \alpha\beta + \beta^2 B = \delta$. 
Therefore, the inner sum is evaluated by
\begin{align}
&\sum^B_{b_1,b_2=1} 
\mathbb{E}
\big[
[\bHtimenoidx_{\ell_1}^\sigma]_{b_2,m}^*
[\bHtimenoidx_{\ell_1}^\sigma]_{b_1,m}
\big]
\mathbb{E}
\big[
[\bHtimenoidx_{\ell_2}^\sigma]_{b_1,n}^*
[\bHtimenoidx_{\ell_2}^\sigma]_{b_2,n}
\big]
\notag\\
&= \frac{B\sigma^4}{W^2} + \frac{\sigma^2 }{LW} (\alpha^2 + \delta)
(\mathbb{I}(\ell_1< L) + \mathbb{I}(\ell_2< L)) 
\notag\\
&+ \frac{1}{BL^2} (\alpha^2 (\alpha^2+2\delta) + B \delta^2) \mathbb{I}(\ell_1< L) \mathbb{I}(\ell_2< L)
\label{eq:0MSE_inntersum}
.
\end{align}
Now, we simplify \fref{eq:0ordererroraproof1} using the results from \fref{eq:0MSE_inntersum} with the fact that $\alpha = \sqrt{1-\delta}$ and $\sum^{W-1}_{\ell_1=0} |{R_{\ell_1 \ell_2}}|^2 = 2W$ so that $\sum^{W-1}_{\ell_1,\ell_2=0} |{R_{\ell_1 \ell_2}}|^2 = 2W^2$. 
Hence,
\begin{align*}
&\MSEzeroIJomega = 
2B \sigma^4
+ 4 \sigma^2
+ 
\frac{1 + \delta^2 (B-1)}{BL^2}\sum^{L-1}_{\substack{\ell_1,\ell_2=0 \\\ell_1\neq\ell_2}}
|{R_{\ell_1 \ell_2}}|^2
\\
&
=
2B \sigma^4
+ 4 \sigma^2 + 
\frac{2(1 + \delta^2(B-1))}{B}
\Big(1 - 
\sum^{L-1}_{\ell=0}
\sum^\ell_{\Delta \ell=-\ell}
\frac{e^{j\theta\Delta \ell} }{L^2}
\Big)\\
&\stackrel{(a)}{=}
2\sigma^2(2 + B\sigma^2) + \frac{2}{B}
(1 + \delta^2(B-1))
(1- f_L(\theta) ),
\end{align*}
where $(a)$ comes from the definition of Fej\'er kernel~\cite{BN1971}. Note that we defined the shorthand variable $\theta = \frac{2\pi}{W}(p_k - \omega)$.} 

\revisiontwo{
The proof can be generalized to per-UE large-scale fading by expressing the channel matrix as $\overline{\bH}_\omega = \bH_\omega \bD$, where $\bH_\omega$ was defined in \fref{eq:Homega_imperfectCSI} and the diagonal matrix~$\bD$ contains the large-scale fading coefficients for the UEs on the main diagonal.
In addition, the proof can be generalized to receive-side correlation matrices~$\bR$ by rewriting $\bHtime$ in \fref{eq:Htime_cor} with 
$\bHtime = \sqrt{\bR}\bHtimeuncor$. A corresponding analysis is left for future work. 
}

\section{Proof of \fref{cor:0th_order_bound}}
\label{app:0orderbound}

\revision{
Since $f_L(\phi)$ is non-negative, it is obvious that $\max_{\omega\in\Omega} \{\MSEzero_\omega \}  \leq \varepsilon_\text{CSI} + 2(1+\varepsilon_\text{cor})/B$  for all $\omega\in\Omega$.}
The equality is satisfied if $(p-\omega)/W = b/L$ for some integer $b>0$ so that $f_L\big(\frac{2\pi}{W}(p-\omega)\big) = f_L\big(\frac{b2\pi}{L}\big) = 0$.
We note that this can only happen if $\dmax\geq W/L$, where $\dmax$ is the maximum distance between any target subcarrier $\omega$ point and its nearest base point; this is due to the fact that $\frac{2\pi}{W}(p-\omega) \leq \frac{2\pi}{W}\dmax <\frac{2\pi}{L}$.

Assume $\dmax<W/L$. Then, by \fref{lem:Fejer1}, the maximum MSE of \zeroth order interpolation is given by:
\begin{align*}
&\max_{\omega\in\Omega} \{\MSEzero_\omega \} 
\\
&= \varepsilon_\text{CSI} + \frac{2}{B}
(1+\varepsilon_\text{cor})
\left(1-\min_{\omega\in\Omega\backslash\setP} f_L\!\left(\frac{2\pi}{W}(p_k-\omega)\right)\right) \notag \\
& = \varepsilon_\text{CSI} + \frac{2}{B}
(1+\varepsilon_\text{cor})
\left(1-f_L\!\left( \frac{2\pi}{W}\dmax\right)\right).
\end{align*}
%
%
%
%
%

\section{Proof of \fref{thm:1ordererror}}
\label{app:1ordererror}

The proof is similar to that of \fref{thm:0ordererror} in \fref{app:0ordererror}. We start by defining the following auxiliary function
\begin{align*}
Q_{\ell \ell^\prime} = & \lambda_\omega 
\exp\!\left(
j\frac {2\pi p_k}{W}\Delta\ell
\right)
+
(1-\lambda_\omega)
\exp\!\left(
j\frac {2\pi p_{k+1}}{W}\Delta\ell
\right)
\\&
- 
\exp\!\left(
j\frac {2\pi \omega}{W}\Delta\ell
\right),
\end{align*}
where we introduced the variable $\Delta\ell = \ell-\ell'$.
\revision{
The result is obtained by substituting $Q_{\ell \ell^\prime}$ in place of $R_{\ell \ell'}$ at \fref{eq:0ordererroraproof1} in \fref{app:0ordererror}.
Note that $\sum^{W-1}_{\ell_1=0} |{Q_{\ell_1 \ell_2}}|^2 = 2(1- \lambda_\omega(1- \lambda_\omega ))W$ which shows that $\sum^{W-1}_{\ell_1,\ell_2=0} |{Q_{\ell_1 \ell_2}}|^2 = 2(1-\lambda_\omega(1- \lambda_\omega ))W^2$.}


\section{Proof of \fref{cor:1orderbound}}
\label{app:1orderbound}

Without loss of generality, we will assume that the target subcarrier index $\omega$ is closer to $p_{k+1}$ so that $\lambda_\omega\in[0,0.5]$.
We will assume that $p_k<\omega<p_{k+1}$ so that $d_k=p_{k+1}-p_k>0$.
Using the results from \fref{app:0ordererror} and \fref{app:1ordererror}, the difference of $\MSEone_\omega$ and $\MSEzero_\omega$ is given by:
\revision{
\begin{align*}
&\MSEone_\omega-\MSEzero_\omega  = -\varepsilon_\text{CSI}  \lambda_\omega (1-\lambda_\omega ))
\\ 
&\quad +  \frac{1+\varepsilon_\text{cor}}{BL^2}\sum^{L-1}_{\ell_1,\ell_2=0}(|{Q_{\ell_1\ell_2}}|^2 - |{R_{\ell_1\ell_2}}|^2 ).
\end{align*}
%
%
Without loss of generality, we assume that $\ell_1-\ell_2>0$ since $|{Q_{\ell_1\ell_2}}|^2 - |{R_{\ell_1\ell_2}}|^2$ is even and is $0$ if $\ell_1=\ell_2$. 
%
We simplify the term $|{Q_{\ell_1\ell_2}}|^2 - |{R_{\ell_1\ell_2}}|^2$ by denoting $\phi=\theta(\ell_1-\ell_2) >\theta$, where $\theta = \frac{2\pi}{W}d_k$ and expand the expression $|{Q_{\ell_1\ell_2}}|^2 - |{R_{\ell_1\ell_2}}|^2$ as follows:
}
\begin{align}
& |{Q_{\ell_1\ell_2}}|^2 - |{R_{\ell_1\ell_2}}|^2 \notag \stackrel{(a)}{=}2\lambda_\omega
\big(
(1-\lambda_\omega)(\cos(\phi)-1)
\\
%
&\quad - (\cos((1-\lambda_\omega)\phi)-\cos(\lambda_\omega\phi))\big) \notag \\
&\stackrel{(b)}{=} 2\lambda_\omega\sin(\phi/2)
(\sin((1-2\lambda_\omega)\phi/2)-(1-\lambda_\omega)\sin(\phi/2)).
\label{eq:diffexpansion}
\end{align}
Here, $(a)$ follows from the definition of ${Q_{\ell_1\ell_2}}$ and ${R_{\ell_1\ell_2}}$ and $(b)$ is a results from simplifying the expression $\cos((1-\lambda_\omega)\phi)-\cos(\lambda_\omega\phi))=-2\sin((1-2\lambda_\omega)\phi/2)\sin(\phi/2)$.

\revision{
We first note that since $\varepsilon_\text{CSI} = 2\sigma^2(2+B\sigma^2)\geq 0$ and by \fref{eq:diffexpansion}, $\MSEone_\omega=\MSEzero_\omega$ if $\varepsilon_\text{CSI}=0$, and $L=1$ or $\lambda_\omega=0$. 
This behavior can be explained intuitively because when $\varepsilon_\text{CSI}=0$ and $L=1$, then the channel is flat across all subcarriers, and hence 
$\MSEone_\omega=\MSEzero_\omega$ 
}
Hence, we now show that $\MSEone_\omega<\MSEzero_\omega$ for $L>1$ and~$\lambda_\omega\neq 0$ by showing that $|{Q_{\ell_1\ell_2}}|^2 - |{R_{\ell_1\ell_2}}|^2<0$. 
First note that $\ell_1-\ell_2<L$ and if $d_k<W/(3L)$, then $\theta<\phi<L\theta = \frac{2\pi}{W}Ld_k < \frac{2\pi}{3}$. Hence, $\sin(\phi/2)>0$.
Therefore, showing that \fref{eq:diffexpansion} is negative is equivalent to:
\begin{align}
 g(\lambda_\omega) = \frac{\sin((1-2\lambda_\omega)\phi/2)}{1-\lambda_\omega}
<\sin(\phi/2).
\label{eq:diffsimplified}
\end{align}
We now prove \fref{eq:diffsimplified} by noting that $g(\lambda_\omega)=\sin(\phi/2)$ if $\lambda_\omega=0$ and $g'(\lambda_\omega)<0$ for all $\lambda_\omega\in(0,0.5]$ so $g(\lambda_\omega)$ is monotonically decreasing in $(0,0.5]$.
The proof is straightforward by:
\begin{align}
g'(\lambda_\omega)  = 
\frac{-\phi\cos((1-2\lambda_\omega)\phi/2)}
{1-\lambda_\omega} + 
\frac{\sin((1-2\lambda_\omega)\phi/2)}{(1-\lambda_\omega)^2},
\label{eq:gderivative}
\end{align}
and, hence, $g'(\lambda_\omega)<0$ in \fref{eq:gderivative} can be expressed as:
\begin{align}
\tan((1-2\lambda_\omega)\phi/2)<(1-\lambda_\omega)\phi.
\label{eq:tan_proof}
\end{align}
To show \fref{eq:tan_proof}, we introduce the shorthand notation $\gamma=1-2\lambda_\omega \in [0,1)$.
With the new notation $\gamma$, the proof is straightforward by:
\begin{align*}
\tan((1-2\lambda_\omega)\phi/2) &\stackrel{(a)}{\leq} \gamma \tan(\phi/2)
\stackrel{(b)}{\leq}  \gamma \phi < (\gamma+\lambda_\omega) \phi \\
&= (1-\lambda_\omega)\phi,
\end{align*}
where $(a)$ follows from the convexity of $\tan(x)$ in $x\in[0,\pi/2)$ and $(b)$ follows from $\tan(\phi/2) < \phi$ for all $\phi\in(0,2\pi/3]$.
Since $g'(\lambda_\omega) <0$ for all $\lambda_\omega\in(0,0.5]$, $g(\lambda_\omega)$ is monotonically decreasing and thus, from~\fref{eq:diffexpansion}, it follows that 
$|{Q_{\ell_1\ell_2}}|^2 - |{R_{\ell_1\ell_2}}|^2 < 0$ for $L>1$.

We conclude by noting that a sharper upper bound on $d_k$ can be obtained by directly computing the bounds for $\MSEone_\omega-\MSEzero_\omega$, i.e., 
\revision{
\begin{align*}
& \MSEone_\omega -\MSEzero_\omega = -\varepsilon_\text{CSI}  \lambda_\omega (1-\lambda_\omega )) + \frac{2\lambda_\omega(1+\varepsilon_\text{cor})}{B}
\\
&\times\Big(
f_L(\lambda_\omega\theta)- 
f_L( (1-\lambda_\omega)\theta) 
-(1-\lambda_\omega) (1-f_L(\theta))
\Big)<0,
\end{align*}
for all $\lambda_\omega\in(0,0.5]$, but we leave an analysis of such refined bounds for future work.
}

\end{appendices}


\balance

\bibliographystyle{IEEEtran} 

\bibliography{bib/VIPabbrv,confs-jrnls,publishers,bib/VIP_181004}

\balance

\end{document}